\documentclass[twocolumn,showpacs,preprintnumbers,amsmath,amssymb,superscriptaddress,10pt,aps,prb]{revtex4-1}
\usepackage{epsfig,amsopn}
\usepackage{graphicx}
\usepackage{epstopdf}
\usepackage{amsmath,amssymb}
\usepackage{amsthm}
\usepackage{enumerate}
\usepackage{xcolor}
\begin{document}

\title{Spin-dependent Andreev reflection in spin-orbit coupled systems by breaking time-reversal symmetry}
\author{Dibya Kanti Mukherjee}
\affiliation{Harish-Chandra Research Institute, HBNI, Chhatnag Road, Jhunsi, Allahabad 211 019, India}
\author{Joanna Hutchinson}
\affiliation{Department of Physics, Indian Institute of Technology Kanpur, Kanpur 208016, India}
\author{Arijit Kundu}
\affiliation{Department of Physics, Indian Institute of Technology Kanpur, Kanpur 208016, India}

\begin{abstract}
  We study theoretically the differential conductance at a junction between a time reversal symmetry broken spin orbit coupled system with a tunable band gap and a superconductor. We look for spin-dependent Andreev reflection (i.e, sub-gap transport) and show that when various mass terms compete in energy, there is substantial difference of Andreev reflection probability depending on the spin of the incident electron. We further analyze the origin of such spin-dependence and show how the incident angle of the electrons controls the spin-dependence of the transport.
  %The dynamics of a spin orbit coupled system, when driven with frequency much larger than the band-width, can be described by an effective static Hamiltonian that is known to have a rich topological phase diagram. 
\end{abstract}

\maketitle
 
\section{Introduction}\label{intro}

Andreev reflection~\cite{Andreev1964} (AR) is a process that occurs at the interface of a normal metal and a superconductor (SC), where an incoming electron from the metallic side is reflected back as a hole. This results in Cooper pair diffusion in the metal and is responsible for a number of observable phenomena, such as proximity induced superconductivity and Josephson current through Andreev bound states, which are formed due to multiple AR processes. As long as the superconductor is of the $s$-wave type, Andreev reflection is suppressed if the metal is ferromagnetic~\cite{Beenakker1995,Cayssol2005} and, in turn, such a suppression of AR is a signature of ferromagnetic exchange in a system~\cite{Upadhyay1998,Soulen1999}.

Haldane model \cite{Haldane1988} is an important model introduced in 1988 to demonstrate quantum-Hall effect in absence of any magnetic field. In Haldane model, time-reversal symmetry (TRS) is broken in 2D systems by introducing complex next-nearest-neighbor (NNN) hopping, which acts like a pseudo spin-orbit (SO) coupling in the orbital space. Although no solid-state realization of such a exists till date, in cold-atomic setup, such a system is, interestingly, realized through periodic drive \cite{Jotzu2014} and also in photonic systems~\cite{Rechtsman2013,Goldman2016} through mechanism that is often called Floquet manipulation of band-structures~\cite{Oka2009, Yao2007, Delplace2013, Katan2013, Lababidi2014, Iadecola2013, Goldman2014, Grushin2014, Dutreix2016, Stepanov2017, Kitagawa2010,Lindner2011,Dora2012,Kundu2014, Titum2015, Klinovaja2016, Thakurathi2013, Chen2016, Khanna2017,Aidelsburger2014}. In this work we consider a similar system with such pseudo SO coupling, along with real SO coupling, which give rise to spin-momentum locking. Representative systems are a number of 2D systems (like silicene, germanene, stanene), typically in a honeycomb lattice, with a small time-reversal (TR) invariant spin-orbit coupling and optionally a sub-lattice staggered potential~\cite{Drummond2012,Liu2011b,Liu2011a,Ezawa2011a,Ezawa2011b}. Andreev reflection, among other transport phenomena, has been studied extensively in these systems~\cite{Linder2014,Li2016,Qiu2016,Wang2015,Chen2013,Li2016,Vosoughi-nia2017,Zhou2016,Li2016,Zhou2017,Dolcini2008} without any pseudo SO coupling. As the spin-symmetry is typically not broken, AR is not suppressed as long as the band gap remains smaller than the superconducting gap. In presence of pseudo SO coupling in the orbital space, the spin-symmetry can now be broken in the band-structure \cite{Ezawa2011a} and we look for possibilities of spin-dependent Andreev reflection.

\begin{figure}
\centering
\includegraphics[width=0.45\textwidth]{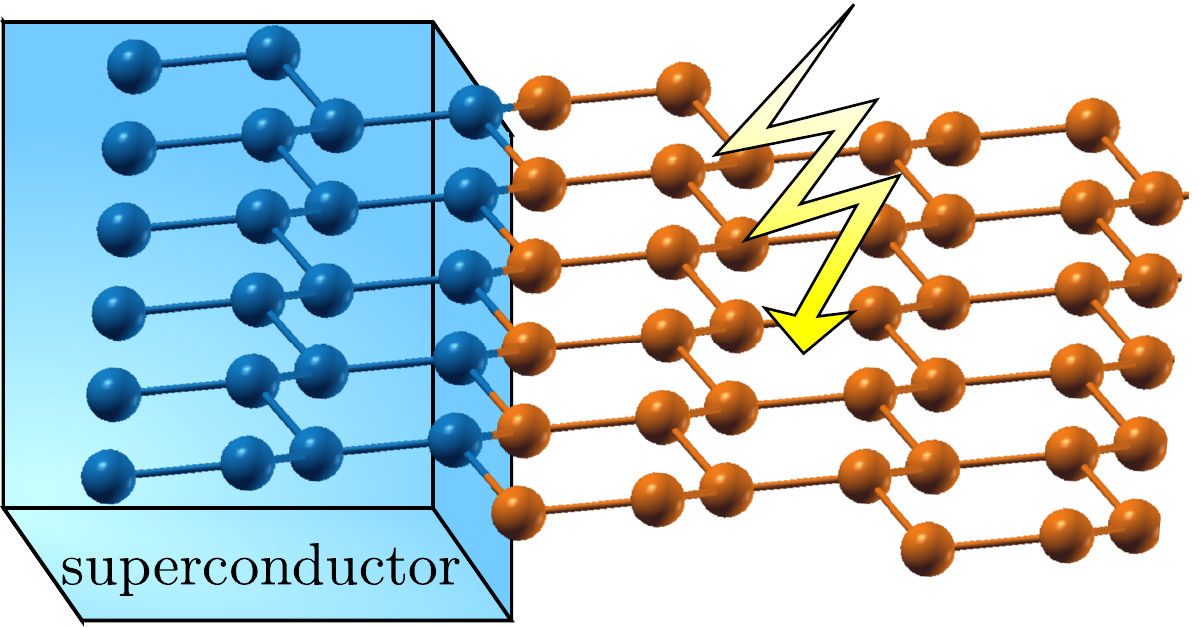} 
\caption{A setup where the model we discuss can be realized. On the right side ($x>0$), the $\mathcal{N}$ region is modeled with spin-orbit coupled two dimensional material (say silicene). Time reversal symmetry is broken with either a complex next-nearest-neighbor hopping term or with circularly polarized light. The left side ($x<0$), the $\mathcal{S}$ is modeled with a proximity induced superconductor of the same material (without the radiation).}\label{fig:setup}
\end{figure}

The setup we consider is shown in Fig.~\ref{fig:setup}, where a part of the system has proximity induced superconductivity (the $\mathcal{S}$ region) and the other part (the $\mathcal{N}$ region) has broken TRS due to complex NNN hopping, representing a system with spin-orbit as well as pseudo spin-orbit coupling. The effective Hamiltonian of the system $\mathcal{N}$, in  presence of TRS breaking, would include an additional mass term ~\cite{Haldane1988}. In this work we explore how the competition between the spin-orbit coupling and the TR breaking mass term can give rise to spin dependent AR reflection probability in certain parameter regimes. 

%------------------------------
%------------------------------

%-------------------------------
\begin{figure}
  \includegraphics[width=0.45\textwidth]{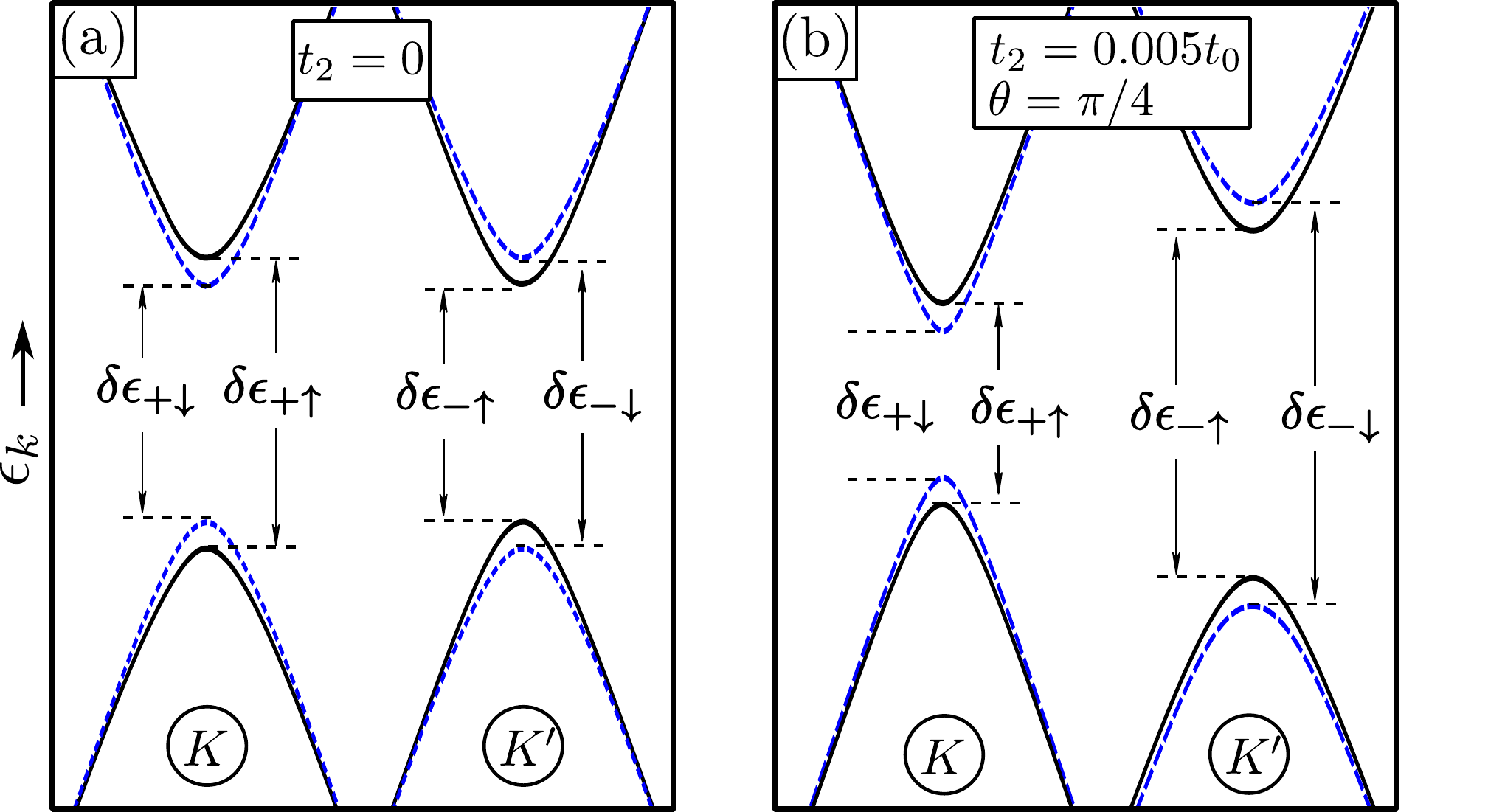}
  \caption{Spin asymmetry in band structure created by next-nearest-neighbor hopping as shown in Eq.~(\ref{Eq:HalLam}). The Hamiltonian of the time reversal symmetric system with $\delta\epsilon_{\eta\sigma}$ = $\delta\epsilon_{-\eta-\sigma}$ is shown in $(a)$ Such symmetry is broken for finite $t_2$ and $\theta$, as shown $(b)$. The parameters used are $lE_z = 0.005t_0$ and $\lambda=0.05t_0$. }
  \label{fig:bands_NNN}
\end{figure}
%-------------------------------

In the case of generating the NNN coupling by periodically driving the system, the $\mathcal{N}$ region in Fig.~\ref{fig:setup}, is irradiated with circularly polarized light with frequency much larger than any other energy scale of the system (such as the natural band-width). Apart from the driving amplitude and the frequency of the drive, it is also possible to control this probability by collimation of the incident angle of the electrons. Further, as the spin-orbit coupling in the relevant system is typically of the order of only a few millielectron-volt~\cite{Liu2011}, the required driving amplitude for the spin polarised conductance to show up is also relatively small.

% \textcolor{red}{rephrase}Among many developments, the effect of the electronic spins has attracted particular attention~\cite{FlSpin1,FlSpin2,FlSpin3}. 

%Even when a system is time dependent in presence of an external driving field, it is often possible to describe the dynamics of the system (ideally at stroboscopic time or when the experimental time scale is much larger than other relevant time scales) by means of a static Hamiltonian~\cite{highf1,highf2,Mikami2016,Mohan}. 

%Although the TR breaking we consider is accomplished by an external driving, this is not necessary for any conclusion of our work. In principle such TR breaking can also be possibly introduced by the orbital effect of a magnetic field, i.e, introducing pseudo-spin-orbit coupling in essence of Haldane model [REF]. We devote a section at the end of the paper to discuss this.

\section{Pseudo Spin-Orbit Coupled system}\label{sec:Haldane}
A two dimensional honeycomb lattice with spin-orbit coupling is represented by the low-energy Hamiltonian
\begin{equation}
\mathcal{H}^{0}_\sigma = 
\begin{pmatrix}
H^0_{+,\sigma} & 0 \\
0 & H^0_{-,\sigma} 
\end{pmatrix}
\end{equation}
where
\begin{equation}\label{eq:h0}
H^{0}_{\eta,\sigma} =  \frac{3t_0a_0}{2}(\eta  k_x \tau_x + k_y \tau_y)+ (lE_z + \eta\sigma\lambda)\tau_z - \mu \mathbb{I}.
\end{equation}
The nearest neighbour hopping, $t_0$, is independent of the spins. $a_0$ is the lattice spacing, $\sigma=\pm$ refers to the up/down spin and $\tau_i$ are Pauli matrices in the sublattice basis. In a buckled structure, the atoms of the sub-lattices are separated in the direction perpendicular to the plane of the lattice. $2l$ is the separation between the $a$ and $b$ sublattices and $E_z$ is the applied electric field. The energy separation $lE_z$ acts as a staggered potential between the sub-lattices. The term $\lambda$ controls the strength of the spin-orbit coupling and $\mu$ is the chemical potential. We note that $\lambda$ only describes the time-reversal (TR) invariant intrinsic spin-orbit interaction and not the Bychkov-Rashba effect, since we expect the latter to be small for such systems~\cite{Liu2011,Konschuh2010}.This description can apply to a variety of topical models such as graphene, silicene, germanene and stanene in honeycomb lattice. The above low-energy Hamiltonian can be derived by expanding the lattice Hamiltonian $H^{\rm hc}$ of such systems near the two inequivalent Dirac points (marked by $\eta=\pm$) in the Brillouin zone.  

%------------------------------
\begin{figure}
	\centering
	\includegraphics[width=0.48\textwidth]{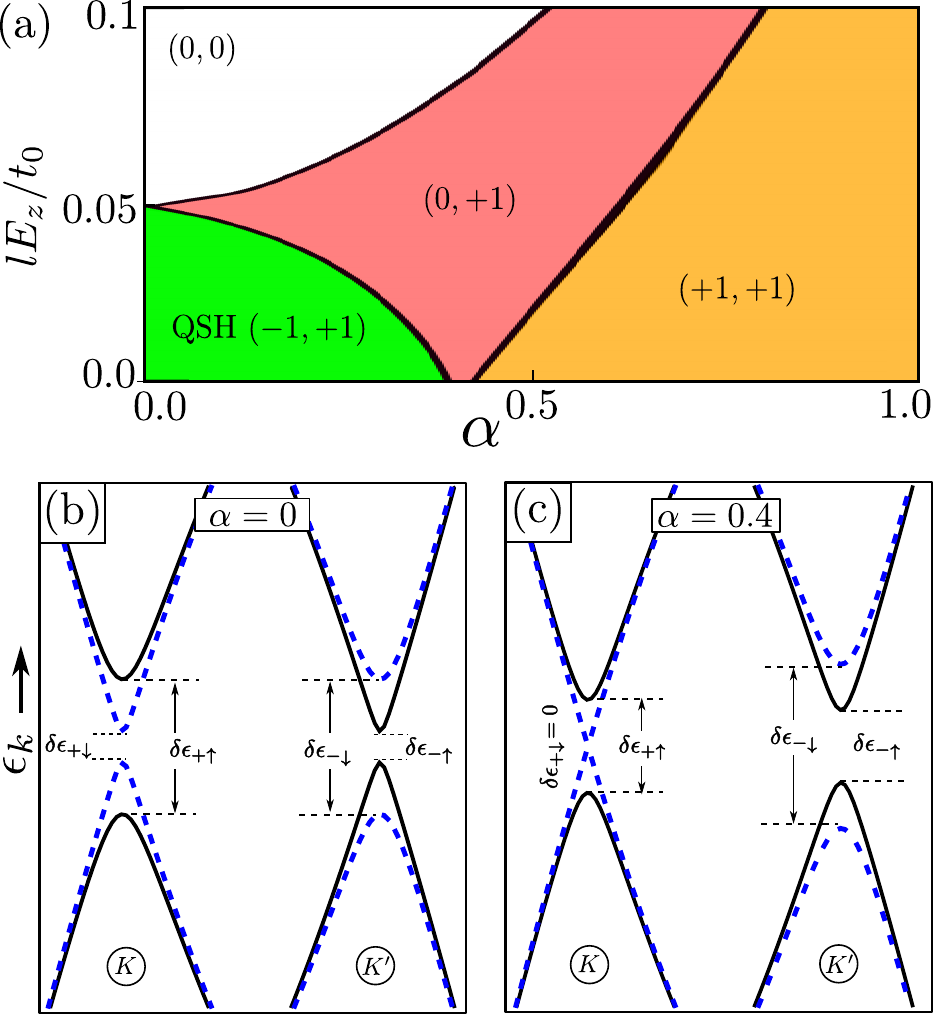} 
	\caption{The topological phases of the system Eq.~(\ref{eq:drivenH}) is shown above in (a), where the chern number of up and down spin bands are shown in parenthesis. The quantum spin-hall system is particularly marked. Periodic drive created spin asymmetry in band structure is shown in (b) and (c). The Hamiltonian of the pristine system ($\alpha=0$), representing silicene is time reversal symmetric giving the relation $\delta \epsilon_{\sigma}^{\eta} = \delta \epsilon_{-\sigma}^{-\eta}$ among the gaps of the system. Such symmetry is broken for finite $\alpha$, as shown in the right. An Andreev reflection process is allowed for an incident electron with energy $E$, spin $\sigma$ and valley index $\eta$, only when $\epsilon\ge \text{max}\{\delta \epsilon_{\sigma \eta},\delta \epsilon_{-\sigma-\eta}\}$. $lE_z=0.08$ for both (a), (b) and $\omega=10t_0$ for (c).}\label{fig:BandCno}
\end{figure}
%------------------------------

%------------------------------
\begin{figure*}
\centering
\includegraphics[width=0.99\textwidth]{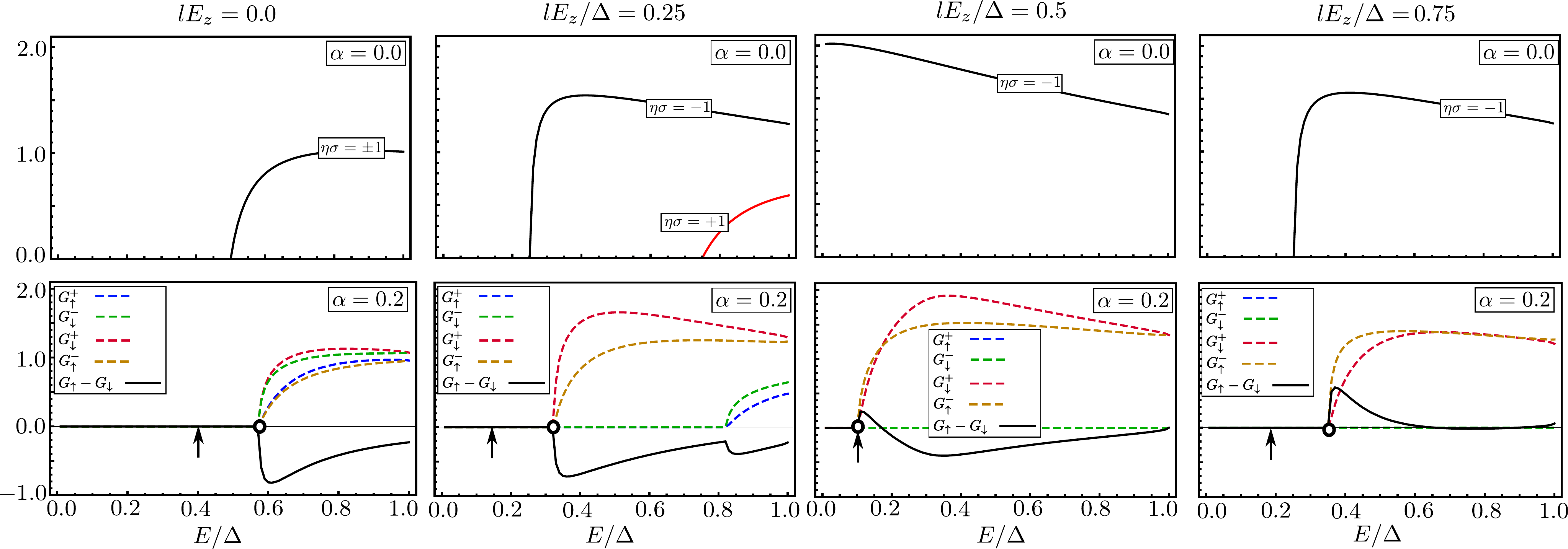} 
\caption{Top panel: The conductances of an electron in the static ($\alpha =0$) system for various values of the staggered potential ($lE_z$). Out of the four channels ($\eta=\pm1, \sigma=\pm1$) the AR probabilities are the same for channels with same value of $\eta\sigma$. For certain values of $lE_z$, only two of the channels with $\eta\sigma=-1$ contribute in transport. Bottom panel: For similar values of $lE_z$, the driven system with $\alpha=0.2$ and $\omega=10t_0$, significant difference in AR probability arises, even when it remains sufficient to consider only two of the channels ($\eta\sigma=-1$). In these figures the arrows on the $E$ axis denote the minimum gaps of the system and the minimum energies $E$ that satisfies Eq.~(\ref{eq:allE}) are marked by circles. The parameters used are $\lambda = 0.5\Delta$, $\Delta = 0.1t_0$.}\label{fig:SPC}
\end{figure*}
%------------------------------

When complex NNN hopping is introduced, the Hamiltonian gets modified\cite{Haldane1988}:
\begin{align}
 H^{\mathcal{N}}_{\text{Hal},\eta,\sigma} =  \frac{3t_{0} a_0  }{2} (\eta  k_x \tau_x + k_y \tau_y)+(l E_z + 3 \sqrt{3} \eta ~\Xi_\sigma)\tau_z - \mu \mathbb{I}.\label{eq:HalH}
\end{align}
Here,
\begin{align}
\Xi_{\sigma} =& \frac{\sigma \lambda}{3\sqrt{3}}-t_2 \sin(\theta),  \label{Eq:HalLam}
\end{align}
where $t_2$ is the NNN hopping amplitude and $+(-)\theta$ is phase associated with the hopping from $A$ to $A$ ($B$ to $B$) sub-lattices. The Hamiltonian results in a similar band-structure
\begin{equation}
\epsilon_{\eta,\sigma}(k)  = 
- \mu  \pm\sqrt{ t^2_0 (k^2_x + k^2_y) + \mathcal{D}^2_{\eta\sigma}},
\end{equation}
with $\mathcal{D}_{\eta\sigma} = lE_z + 3\sqrt{3}\Xi_{\sigma}$ is the gap introduced by the TRS breaking. Fig.~\ref{fig:bands_NNN} describes the band structure before and after the time reversal symmetry is broken by introducing the NNN hopping term.

In Fig.~\ref{fig:setup}, the $\mathcal{N}$ region,$x>0$, is described by the Eq.~(\ref{eq:HalH}). On the other hand, the region $\mathcal{S}$ is modeled by the system along with proximity induced $s$-wave superconductivity. For $x<0$ (region $\mathcal{S}$) we take the Hamiltonian Eq.~(\ref{eq:h0}) and set $lE_z=0$ for simplicity. Further we need to take a large doping $U_0$ for the mean-field description of the superconducting part to remain valid. The pair potential $\Delta$ (which we consider to be real) couples the time-reversed electron and hole states in the superconductor. Hence, we arrive at the low-energy Hamiltonian of the $\mathcal{S}$ side \cite{Linder2014}: 
\begin{equation}
H^{SC}_{\eta,\sigma} =\begin{pmatrix}
 H^0_{\eta,\sigma} & \sigma \Delta \mathbb{I} \\
\sigma \Delta  \mathbb{I} & - H^{0}_{\eta,\sigma} 
\end{pmatrix}.
\end{equation}
where $H^0_{\eta,\sigma}$ is the Hamiltonian of the static system in Eq.~(\ref{eq:h0}). For our numerical simulations we have used $U_0=2t_0$.

Either $\theta$ or $t_2$ can be varied to control $\mathcal{D}_{\eta\sigma}$. This gap will compete with the real spin-orbit coupling energy scale $\lambda$ and in suitable situation spin-dependent Andreev reflection might be observed.

\section{Time Reversal symmetry breaking by Periodic drive}\label{model}
As a physical realization of the model discussed in the previous section, we next turn our attention to a similar two dimensional systems with spin orbit coupling. Time reversal symmetry in this system is now broken when circularly polarized light of frequency $\omega$ is irradiated on it. This high frequency drive is represented by a time dependent vector potential $A(t) = A_0(\cos \omega t, \sin\omega t,0)$. 

The irradiation can be treated in the perturbative high-frequency approximation where $\omega$ is the largest energy scale of the system. In the presence of the radiation, the hopping elements of the honeycomb lattice Hamiltonian of Eq.~(\ref{eq:h0}), $H^{\rm{hc}}$, are modified by Peierls substitution, making the Hamiltonian $H^{\rm{hc}}(t)$ time periodic with the $n$th Fourier component being $H^{\rm{hc}(n)}$. In the high-frequency limit, the effective Hamiltonian that controls the dynamics of the system is given by $H^{\rm{hc}}_{\rm{eff}} \approx H^{\rm{hc}(0)}+ \sum_{n\ne0}H^{\rm{hc}(-n)}H^{\rm{hc}(n)}/n\omega$. Expanding this effective Hamiltonian near the Dirac points provides us with a low-energy Hamiltonian for the $\mathcal{N}$ region:
\begin{equation}\label{statham}
\mathcal{H}^{\mathcal{N}}_\sigma = 
\begin{pmatrix}
H^{\mathcal{N}}_{+,\sigma} & 0 \\
0 & H^{\mathcal{N}}_{-,\sigma} 
\end{pmatrix}
\end{equation}
where~\cite{Mohan2016}
\begin{equation}
H^{\mathcal{N}}_{\eta,\sigma} =  \frac{3t_{\sigma} a_0  }{2} (\eta  k_x \tau_x + k_y \tau_y)+(l E_z + 3 \sqrt{3} \eta \Lambda^0_\sigma)\tau_z - \mu \mathbb{I},\label{eq:drivenH}
\end{equation}
with
 \begin{align}
 t_{\sigma} =& t_0 J_0(\alpha)-\frac{4t\sigma \lambda }{3\omega}\sum_{n\ne0}\frac{J_n(\alpha) J_n (\alpha \sqrt{3})}{ \sqrt{3}n}\times\nonumber \\
& \times \left(2\sin{\frac{n\pi}{2}}+\sin{\frac{\pi n}{6}}\left(1+\frac{1}{2}(-1)^n\right)\right), \label{Eq:J}
\end{align}
 \begin{align}
\Lambda^0_{\sigma} =& \frac{\sigma \lambda J_0( \alpha \sqrt{3})}{3\sqrt{3}}-\sum_{n\ne0} \frac{t_0^2J^2_n(\alpha)}{\omega n}\sin{\frac{2\pi n}{3}},  \label{Eq:Lam}
\end{align}
where $\alpha=a_0A_0$ characterizes the strength of the drive and $J_n$ is the Bessel function of order $n$. We have $t_\sigma \sim t_0$ for small $\alpha$ and large $\omega$. It is important to note that the spin symmetry of the static system (i.e, the energy-dispersion of Eq.~(\ref{eq:h0}) remaining the same under transformation $\sigma,\eta \rightarrow -\sigma,-\eta$) is now broken in the presence of the periodic driving. The polarization of the time dependent field breaks the TR symmetry of the Hamiltonian. The energy spectrum is thus given by
\begin{equation}
\epsilon_{\eta,\sigma}(k)  = 
- \mu  \pm\sqrt{ t^2_\sigma (k^2_x + k^2_y) + \mathcal{D}^2_{\eta\sigma}},
\end{equation}
with the redefined mass term $\mathcal{D}_{\eta\sigma} = lE_z + 3\sqrt{3}\Lambda_{\sigma}^0$.

From this it is evident that in the region $\mathcal{N}$ we have a tunable gap between the conduction and valence band given by $\delta \epsilon_{\eta\sigma} = 2 | \mathcal{D}_{\eta\sigma}|$. The presence of three energy scales: the spin-orbit coupling, the staggered potential and the TR breaking mass from the driving gives rise to a rich topological phase diagram, where one can have trivial insulating, Chern insulating as well as spin hall insulating states, with topological phase transitions separating one phase from another~\cite{Mohan2016}.  In Fig.~\ref{fig:BandCno} we briefly summarize these points. It is also known that in this situation it is possible to achieve purely spin polarized low-energy band-structures~\cite{Ezawa2011a}, that would result in the suppression of AR.

%------------------------------
\begin{figure}
\centering
\includegraphics[width=0.45\textwidth]{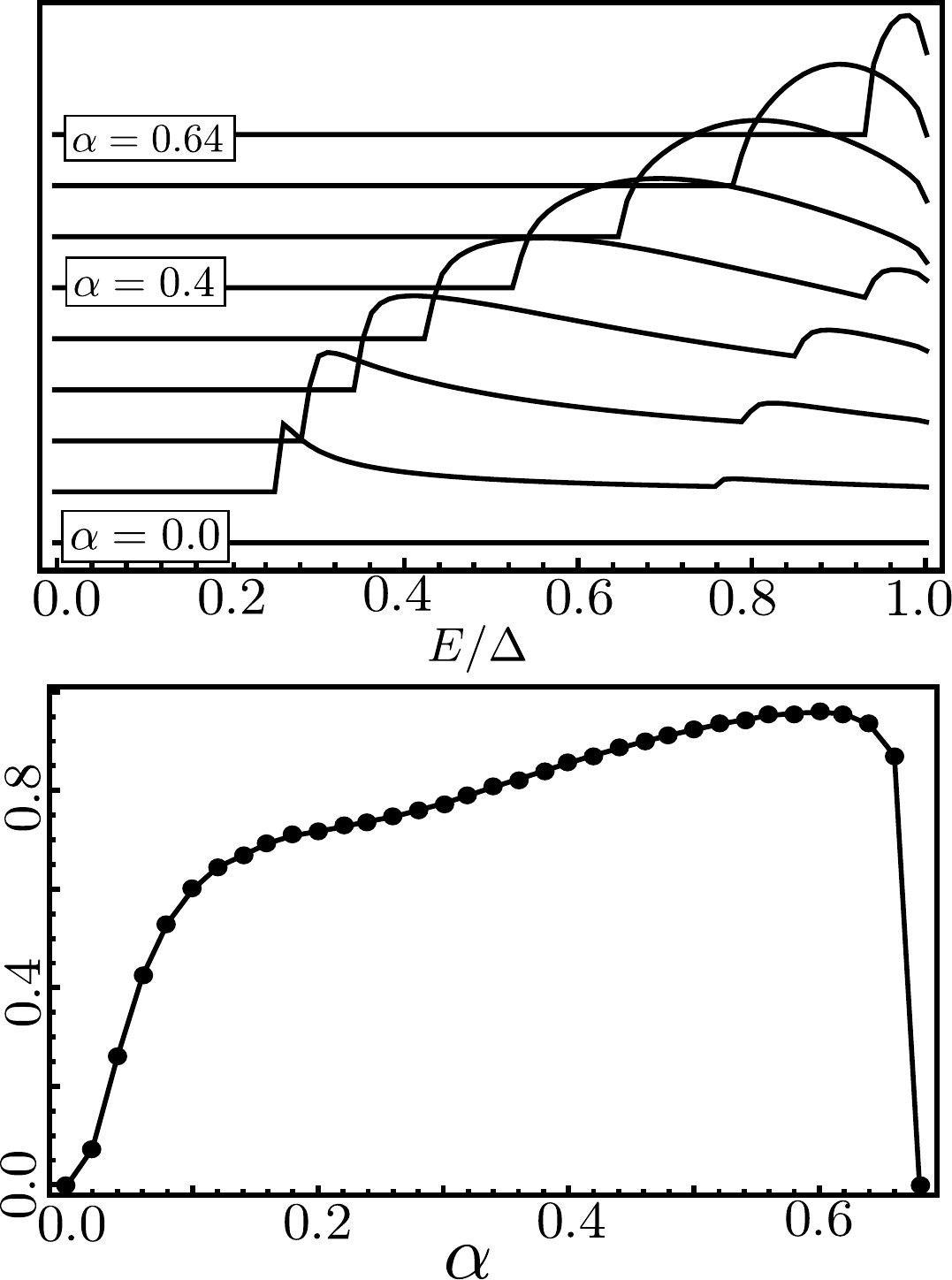} 
\caption{The amplitude of the difference in conductances $\sum_{\eta}(G_{\eta\downarrow} - G_{\eta\uparrow})$ plotted against $E/\Delta$. As $\alpha$ increases the system develops a gap at the $K$ and $K'$ points and consequently with larger $\alpha$, the sub-gap conductance vanishes, giving rise to zero spin-dependent value. For the choice of our parameters: $lE_z = 0.25\Delta$, $\lambda = 0.5\Delta$ and $\Delta = 0.1t_0$, the critical value of $\alpha$ is $~$0.64. The value of the driving frequency is chosen to be $\omega=10t_0$. Surprisingly, the maximun value of the spin-dependent conductance, as a function of $E/\Delta$ ($E$ being the incident energy) remains partly constant with $\alpha$, shown in the second figure.}\label{fig:max}
\end{figure}
%------------------------------

\section{Scattering matrix formalism}\label{chap:ScatMat}
The quantum states can be found in both $\mathcal{N}$ and $\mathcal{S}$ region by solving the BdG equations in the respective regions. To compute the probability of Andreev reflection we match the wavefunctions for regions $\mathcal{N}$ and $\mathcal{S}$ at the boundary $x=0$ in familiar fashion~\cite{Beenakker2006}:
\begin{equation}
\Psi^{e^-} + r \Psi^{e^+} + r_A \Psi^{h^+} = b \Psi^{S^+} + d \Psi^{S^-}
\end{equation}
where $\Psi^{e^-}$ and $\Psi^{e^+}$ are the wave functions of the incident and reflected electron (in band $\eta,\sigma$). $\Psi^{h^+}$ is the wavefunction of the reflected hole (in band $-\eta,-\sigma$):
\begin{align}
\Psi^{e^\mp} =  & \frac{e^{i (\mp k_e x + k_y y)}}{\sqrt{\cos \phi_i}}
\left(
\mp \frac{\xi_{\eta,\sigma}^{\frac{1}{4}}}{\theta_{\eta,\sigma}^{\frac{1}{4}}} e^{ \pm \frac{ i \phi_i}{2}},\eta\frac{\theta_{\eta,\sigma}^{\frac{1}{4}}}{\xi_{\eta,\sigma}^{\frac{1}{4}}} e^{ \mp\frac{ i \phi_i}{2}},0,0\right)^T, \nonumber \\
\Psi^{h^+} = &\frac{e^{i (k_h x + k_y y)}}{\sqrt{\cos \phi'}}
\left(0,0,- \frac{\theta_{-\eta,-\sigma}^{\frac{1}{4}}}{\xi_{-\eta,-\sigma}^{\frac{1}{4}}} e^{ -\frac{ i \phi'}{2}},\eta\frac{\xi_{-\eta,-\sigma}^{\frac{1}{4}}}{\theta_{-\eta,-\sigma}^{\frac{1}{4}}} e^{ \frac{ i \phi'}{2}}\right)^T,\nonumber
\end{align}
where
\begin{align}
  & \xi_{\eta,\sigma} =    \epsilon_{\eta,\sigma} + \mathcal{D}_{\eta\sigma}, \quad \theta_{\eta,\sigma} = \epsilon_{\eta,\sigma} - \mathcal{D}_{\eta\sigma}.\nonumber
\end{align}

%------------------------------
\begin{figure*}
\centering
\includegraphics[width=0.99\textwidth]{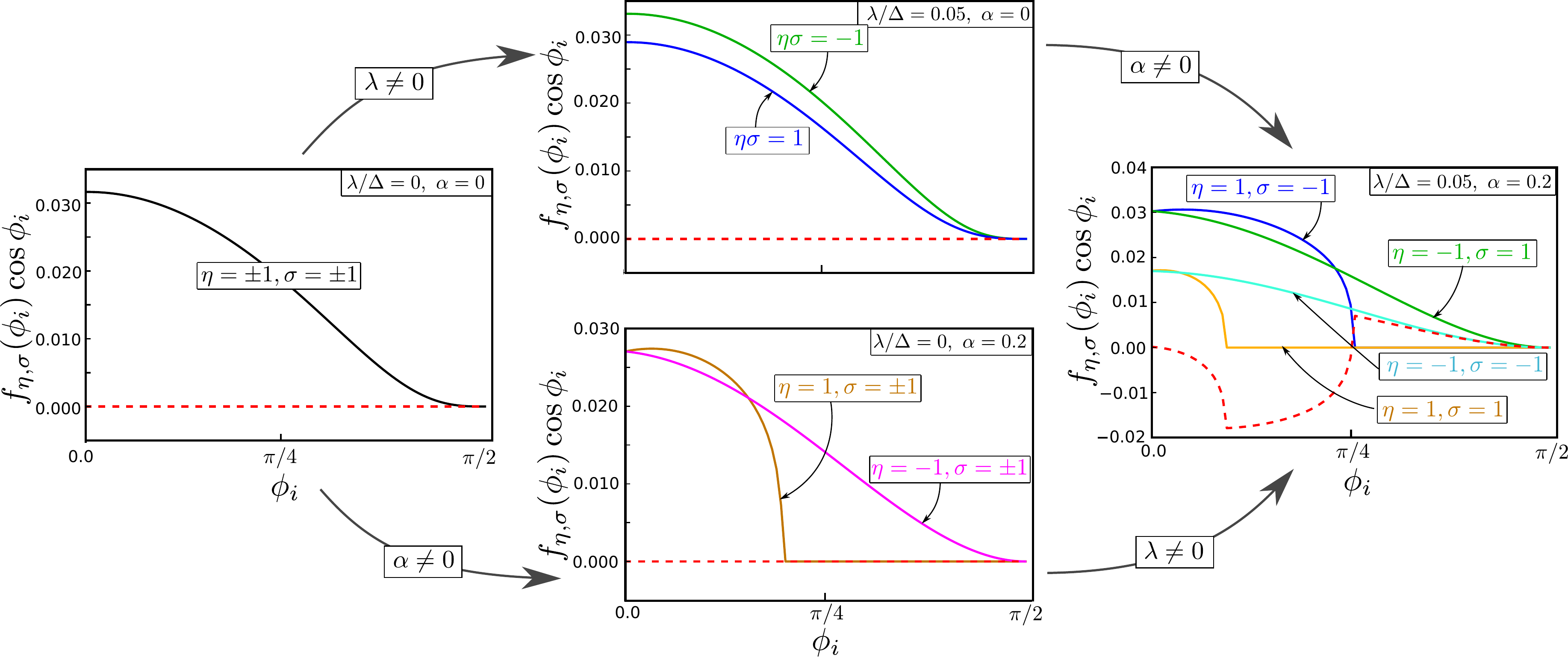} 
\caption{Appearance of spin-dependent AR in presence of both the spin-orbit coupling ($\lambda\neq0$) and periodic driving ($\alpha\neq0$). We plot the angle resolved sub-gap ($E/\Delta=0.4$) conductance $f_{\eta,\sigma}(\phi_i)$ (c.f. Eq.~(\ref{eq:cond})) as a function of the incident angle $\phi_i$ for all the channels available ($\eta=\pm1,~\sigma=\pm1$), as marked individually. The resultant spin-current is plotted in dashed (red) line. Non-vanishing $\lambda$ prefers two (given by $\eta\sigma=-1$) out of the four channels. Whereas, non-vanishing $\alpha$ prefers one valley (here, $\eta=+1$). In combination of both present, we observe spin-dependent AR. $lE_z=0.25\Delta$, $\omega=10t_0$ and $\Delta/t_0=0.1$ are kept fixed.}\label{fig:withLam}
\end{figure*}
%------------------------------

$\phi_i$ is the angle of incidence of the electron and $\phi'$ is the angle of the reflected hole given by
\begin{align}
\cos \phi_i =& \frac{t_ {\sigma} k_e}{\sqrt{\xi_{\eta,\sigma}}\sqrt{\theta_{\eta,\sigma}}}, ~ \tan\phi_i = \eta k_y/k_e,\nonumber \\
\cos \phi'=&\frac{t_{-\sigma} k_h}{\sqrt{\xi_{-\eta,-\sigma}}\sqrt{\theta_{-\eta,-\sigma}}},~ \tan\phi' = \eta k_y/k_h.\nonumber
\end{align}
Here we have set $\mu=0$. The incident angle $\phi_i$ has an upper limit for Andreev reflection to take place. This \textit{critical angle} is given by 
\begin{align}
\phi_c = \sin^{-1}\frac{t_{\sigma}\sqrt{\xi_{-\eta,-\sigma}}\sqrt{\theta_{-\eta,-\sigma}}}{t_{-\sigma}\sqrt{\xi_{\eta,\sigma}}\sqrt{\theta_{\eta,\sigma}}}.
\end{align}
In the $\mathcal{S}$ region, the relevant wavefunctions are
\begin{align}
\Psi^{S^\pm} = & e^{i((\pm k_0 - i \kappa) x+ k_y y)} 
\begin{pmatrix}
e^{\mp i \beta} \\
\pm \eta r_{\pm} e^{i (\pm\gamma_{\pm} \mp \beta) }\\
1 \\
\pm \eta r_{\pm} e^{\pm i\gamma_{\pm}}
\end{pmatrix}\nonumber
\end{align}
with
\begin{align}
& \sin \gamma^{\eta,\sigma}_\pm = \frac{ (3/2) a_0 \eta t_0 k_y}{ \sqrt{\eta \sigma \lambda - U_0 \mp i  Q} \sqrt{-\eta \sigma \lambda-U_0 \mp i  Q}}, \nonumber \\
& r^{\eta,\sigma}_\pm = \sqrt{\frac{- \eta \sigma \lambda -U_0 \mp i Q}{\eta \sigma \lambda -U_0 \mp i Q}}, \quad Q = \sqrt{|\Delta|^2 - \epsilon^2_{\eta,\sigma}}, \nonumber \\
& k_0 = \frac{2 \sqrt{M}}{3 a_0 \eta t_0 }, \quad \kappa= \frac{2 U_0 Q }{3a_0 \eta t_0\sqrt{M} }, \nonumber\\
& M = U_0^2 -Q^2 -\left( \eta \sigma \lambda \right)^2 - \left(\frac{ 3 a_0 \eta t_0 k_y}{2}\right)^2.\nonumber
\end{align}
$\beta$ is the phase associated with Andreev reflection and $\beta = \cos^{-1} (\epsilon/|\Delta|)$ for $\epsilon < \Delta$ and $\beta = -i \cosh^{-1} (\epsilon/|\Delta|)$ for $\epsilon > \Delta$. The solved parameters $r$ and $r_A$, for each of the band, are elements of the scattering matrix of the system, where the probability of reflection and Andreev reflection are, respectively, $|r|^2$ and $|r_A|^2$. Finally, the differential conductance at the $\mathcal{NS}$ junction is given by the Blonder-Tinkham-Klapwijk formula
\begin{equation}\label{eq:cond}
G_{\eta,\sigma} =  \int^{\frac{\pi}{2}}_0f_{\eta,\sigma}(\phi_i) \cos \phi_i~d \phi_i,
\end{equation}
where $f_{\eta,\sigma}(\phi_i) =\left( 1- |r|^2 + |r_A|^2 \right)$ for each incident channel. For sub-gap conductance, as in our case, one can equivalently write $f_{\eta,\sigma}(\phi_i) \equiv 2|r_A|^2 $, as $|r|^2 + |r_A|^2=1$ for each channel. Also, we note that $G$ is measured with respect to the ballistic conductance of the $\mathcal{N}$ system in absence of the superconductor.

Lastly, we would like to point out that under redefinition of parameters, the two Hamiltonians given in Eq.~(\ref{eq:HalH}) and Eq.~(\ref{eq:drivenH}) are the same. So, the formalism of this section can be easily used to study the differential conductance in the geomwtries described in both Sec~(\ref{sec:Haldane}) and (\ref{model}).

\section{Spin-dependent Andreev reflection}
For $\mu=0$, in order for Andreev reflection to occur, the excitation gap in region $\mathcal{N}$ must be smaller than the superconducting gap $\Delta$. Thus, at $\alpha =0$ and for a large-enough $lE_z \sim\mathcal{O}(\Delta)$, it is enough to consider only one pair of bands ($\eta\sigma=-1$) to participate in Andreev processes~\cite{Linder2014} and for incident energy larger than the gap of this pair of bands one expects AR to occur. Such a simplification is not possible for finite $\alpha$ as the degeneracy among the bands is now lifted. An electron coming in the band of $\eta,\sigma$ Andreev reflects to the band with indices $-\eta,-\sigma$. This provides us with a condition that the AR is allowed only when 
\begin{align}
&\Delta\ge E\ge \text{max}\{\delta \epsilon_{\sigma \eta},\delta \epsilon_{-\sigma-\eta}\}. \label{eq:allE}
\end{align}

%------------------------------
\begin{figure}[ht]
\centering
\includegraphics[width=0.4\textwidth]{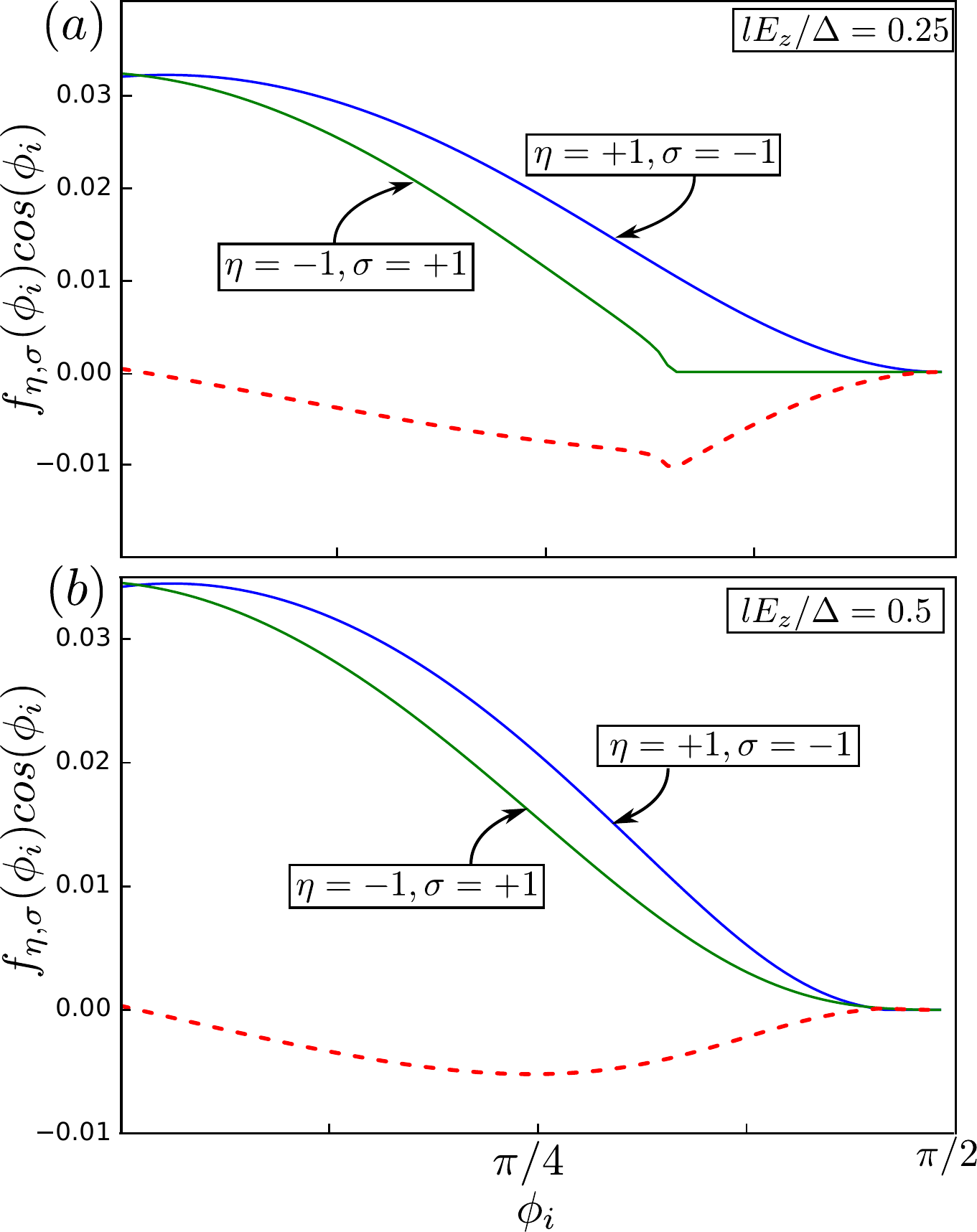} 
\caption{(a) The angle resolved sub-gap conductance $f_{\eta,\sigma}(\phi_i)$ is plotted as a function of the incident angle $\phi_i$ and the resultant spin-resolved conductance is plotted with dashed lines. Only $\eta\sigma=-1$ branches contribute for our choice of parameters: $E = 0.6\Delta$, $\lambda = 0.5\Delta$, $\Delta = 0.1 t_0$, $lE_Z = 0.25\Delta$. (b) Same plot for $lE_z = 0.5\Delta$. Other parameter values remain unchanged from Fig.~\ref{fig:SPC}.}\label{fig:arcond}
\end{figure}
%------------------------------

As an example, for the case considered in Fig.~\ref{fig:BandCno}, for $\alpha=0.4$, although the band $\sigma=-1,\eta=+1$ is gapless (i.e, $\delta\epsilon_{+\downarrow}=0$), the AR takes place only when $E\ge \delta\epsilon_{-\uparrow}$. This simply implies that in a purely spin-polarized band-structure AR is prohibited (i.e, in a range of energy from $\text{min}\{\delta \epsilon_{\sigma \eta},\delta \epsilon_{-\sigma-\eta}\}$ to $\text{max}\{\delta \epsilon_{\sigma \eta},\delta \epsilon_{-\sigma-\eta}\}$). Any occurrence of spin-dependent AR is still not evident yet. However, this does not prohibit the probabilities of the AR in various channels to differ from each other as long as $E$ satisfies Eq.~(\ref{eq:allE}).

Before we discuss the Haldane model, we first study the case of the driven system. We start by briefly summarizing the results of the static system in the upper panel of Fig.~\ref{fig:SPC}. Introduction of the spin-orbit coupling term $\lambda$ breaks the four-fold degeneracy of the Dirac points, but keeps the band-structure symmetric with respect to $\eta,\sigma \rightarrow -\eta, -\sigma$.  Further, due to the presence of the sub-lattice staggered potential $lE_z$, the two branches $\eta\sigma=\pm1$ are now separated by a gap. Consequently, for $lE_z \sim \lambda$, it becomes sufficient to consider only one of the $\eta\sigma$ branches for low-energy transport. The $\lambda$ term in Eq.~(\ref{eq:h0}) does not break the TR symmetry and for $lE_z<\lambda$, the system is a spin-hall insulator with opposite Chern numbers for up and down-spin valence bands. For $lE_z>\lambda$ the system becomes a trivial insulator. As the spin-valley symmetry remains intact, the AR probability remains independent of spin. One can compare these results with that of Ref.~\cite{Linder2014}.

%This is required such that an incoming electron with energy $E>E_g/2$ will have energy less than $\Delta/2$, and thus can be normally reflected within the same condition band or Andreev reflected. In the latter case, the incoming electron is paired with another electron inside the superconductor to join the condensate of cooper pairs, emitting a hole back into region N. There are 2 types of Andreev reflection that can occur \cite{beenakker},
%\begin{itemize}
%\item specular, which dominates for $E_f << \Delta, \epsilon$ associated with inter-band electron-hole conversion.
%\item retro, where the hole is reflected back along the path of the incident electron, which dominates for $E_f >>\Delta, \epsilon$
%\end{itemize}

%-------------------------
\begin{figure}
  \includegraphics[width=0.5\textwidth]{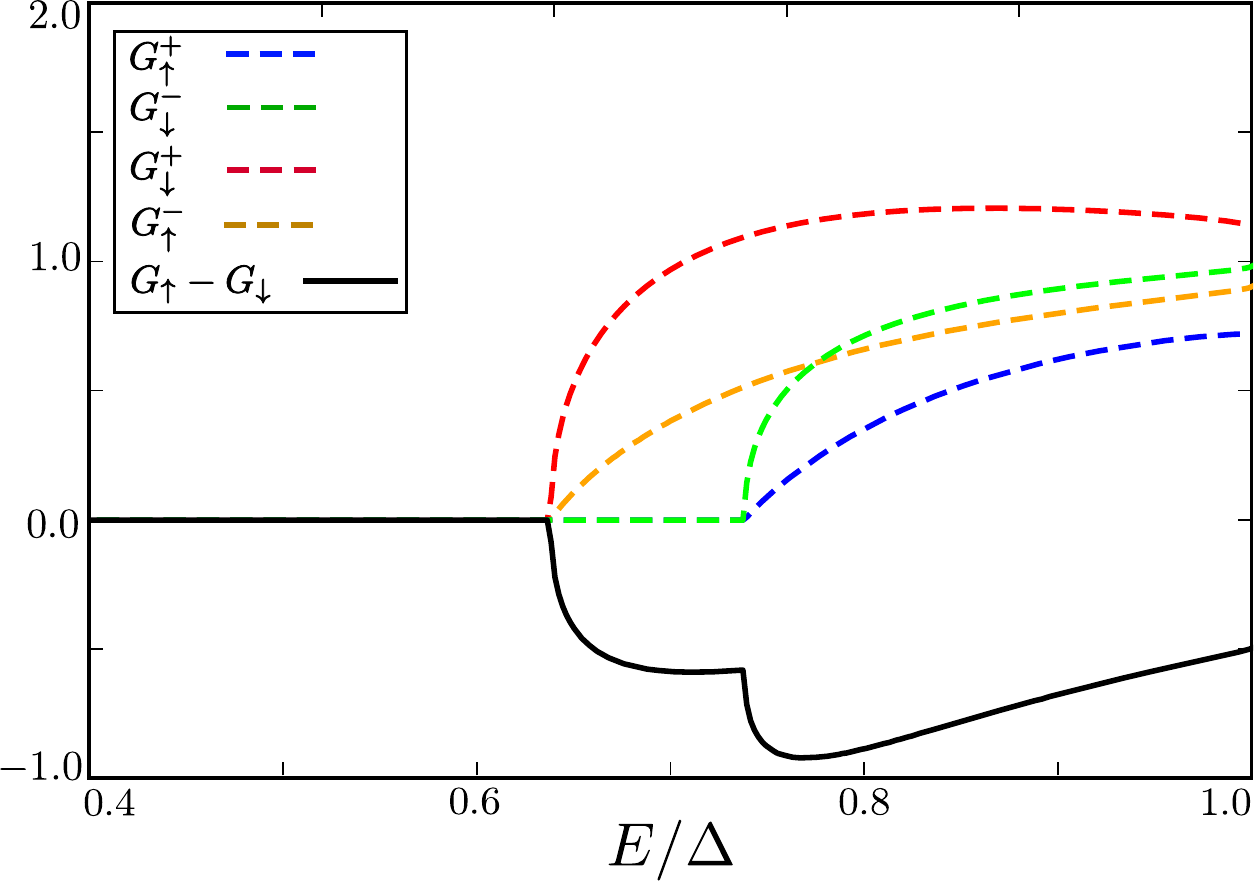}
  \caption{The conductances of an electron when the TRS is broken by complex NNN hoppings as is shown in Eq.~(\ref{eq:HalH}). The parameters taken are $t_2=0.005t_0$, $\theta=\pi/4$, $lE_z=0.005t_0$, $\Delta = 0.1t_0$, $\lambda=0.05t_0$.}
  \label{fig:pseudospin}
\end{figure}
%-------------------------

Breaking the TR symmetry by introducing the driving (characterized by the amplitude $\alpha$) has dramatic consequence in Andreev reflection. We summarize the results depicting spin-dependent AR probabilities in the lower panel of Fig.~\ref{fig:SPC} and in Fig.~\ref{fig:max}. In the presence of $\alpha\neq0$, the four bands ($\sigma=\pm1,~\eta=\pm1$) are now split and there exists a range of energy that does not satisfy Eq.~(\ref{eq:allE}). When the energy $E$ is larger than this forbidden energy, each of the previously equivalent spin channels labeled by the value of $\eta\sigma$ acquires different probabilities of Andreev reflection. Consequently the conductance $G$ becomes spin-dependent. Even for $lE_z=0$, the four channels $\eta=\pm1,~\sigma=\pm1$ are now split and a significant spin-conductance can be observed. With increasing $lE_z$, only two channels, given by $\eta\sigma=-1$ remain relevant in the subgap regime, which continue to carry large spin conductance. For our numerical results, we keep $\omega=10t_0$, which is almost double of the band-width of the system. This is well within the regime where the high-frequency approximation is expected to be valid. 

Our results show spin-dependent conductance even for a comparatively small value of $\alpha/t_0$. This is because a finite spin-dependent AR appears due to the competition between the two terms in Eq.~(\ref{Eq:Lam}) (see the discussion below). The second term appearing in $\Lambda_{\sigma}$ (see Eq.~(\ref{Eq:Lam})) needs to be of the order of $\lambda$ (the spin-orbit coupling strength) for this spin dependence to show up. In our simulation we have taken $\lambda = 0.05t_0$. In typical systems, $\lambda$ is quite small, for example, the value of $\lambda$ in silicene is only $3.9$ meV whereas $t_0\approx1.6$ eV~\cite{Liu2011}, giving a value $\lambda/t_0 \approx 2.5\times 10^{-3}$. We expect the relevant value of $\alpha/t_0$ to be of the same order for such a system.

With increasing $\alpha$, the system eventually develops a gap (a topological insulator with Chern number $=\pm2$, see Fig.~\ref{fig:BandCno}) bigger than $\Delta$ and consequently the sub-gap conductance vanishes. Interestingly, the maximum spin-conductance observed depends only weakly on $\alpha$. These results are summarized in Fig.~\ref{fig:max}.

Now we turn to further analysis of the origin of spin-dependent AR in our system. For that, we study how the angle dependent differential conductance $f_{\eta,\sigma}(\phi_i)$ behaves as a function of the incident angle $\phi_i$. As shown in Fig.~\ref{fig:withLam}, a finite value of $\lambda$ and $\alpha$ is achieved via two different intermediate states. The action of a finite $\lambda$ prefers two channels ($\eta\sigma=-1$ for this case) out of four. Whereas, a finite $\alpha$ results in a critical angle $\phi_c<\pi/2$ for channels belonging to one of the valley ($\eta=1$), giving rise to net conductance dominated by channels of the other valley ($\eta =-1$). Now, in the $\eta\sigma=-1$ channel consisting $\eta=1,\sigma=-1$ and $\eta=-1,\sigma=1$, the effect of driving, for the parameter range presented in Fig.~\ref{fig:withLam}, is a finite critical angle of the $\eta=1,\sigma=-1$ channel, making the net conductance spin-dependent. Thus, when both the effects are present, it becomes clear that the sub-gap conductance has spin-dependence. The angle resolved conductance study also hints at achieving a purely spin polarized sub-gap transport at a range of collimation angle of the incident electron. The angle-resolved sub-gap conductance in the parameter range of Fig.~\ref{fig:SPC} is presented in Fig.~\ref{fig:arcond} for a sample value of the incident energy.

In case of generating the mass in a periodically driven setup, we neglect any other degrees of freedom present in the system (such as phonon and the effect of the substrate), which may limit our prediction for a realistic setup. The heating, when driving with frequency larger than the band width is likely to be negligible~\cite{Polkovnikov2015,Maricq1982,Mitra2014,Mori2016} but may require appropriate cooling of the substrate. Despite the advantage of periodic drive in terms of the tunability of the time-reversal broken mass term, the time dependent drive has significant limitation in solid state systems, where a periodic drive, in presence of interaction, can heat the system eventually.

Lastly, the same study is repeated for the Haldane model introduced in Sec.~\ref{sec:Haldane}. The results are summarized in Fig.~\ref{fig:pseudospin}. As expected, the spin asymmetry due to the next-nearest-neighbor hopping, results in spin-dependent Andreev reflection amplitudes which can be tuned by $t_2$ and $\theta$.

\section{Summary}
To summarize, we consider a simple Dirac system in the presence of a number of mass terms that may compete with one another. In such a system the Andreev reflection probability, hence the sub-gap conductance, at the interface with a superconductor becomes spin-dependent. It is possible to achieve, as we see in Fig.~\ref{fig:withLam}, sub-gap spin dependent transport. If we consider the case where the time-reversal broken mass term is introduced by circularly polarized light, it is possible to control the spin-dependence, by controlling the amplitude of drive $\alpha$, frequency $\omega$ and the handedness of the radiation. Our work is a proof of concept how spin-dependent AR can be achieved and has direct implication in practical systems like silicene, germanene and stanene, as well as in cold-atomic setup.

\section{Acknowledgement}
J. H. gratefully acknowledges support from the Leverhulme Trust during this work. A. K. acknowledges the funding support from the Indian Institute of Technology - Kanpur. A. K. and D. K. M. acknowledges useful discussion with Sumathi Rao. The research of D. K. M. was supported in part by the INFOSYS scholarship for senior students.

\bibliography{bib}

\begin{thebibliography}{51}%
\makeatletter
\providecommand \@ifxundefined [1]{%
 \@ifx{#1\undefined}
}%
\providecommand \@ifnum [1]{%
 \ifnum #1\expandafter \@firstoftwo
 \else \expandafter \@secondoftwo
 \fi
}%
\providecommand \@ifx [1]{%
 \ifx #1\expandafter \@firstoftwo
 \else \expandafter \@secondoftwo
 \fi
}%
\providecommand \natexlab [1]{#1}%
\providecommand \enquote  [1]{``#1''}%
\providecommand \bibnamefont  [1]{#1}%
\providecommand \bibfnamefont [1]{#1}%
\providecommand \citenamefont [1]{#1}%
\providecommand \href@noop [0]{\@secondoftwo}%
\providecommand \href [0]{\begingroup \@sanitize@url \@href}%
\providecommand \@href[1]{\@@startlink{#1}\@@href}%
\providecommand \@@href[1]{\endgroup#1\@@endlink}%
\providecommand \@sanitize@url [0]{\catcode `\\12\catcode `\$12\catcode
  `\&12\catcode `\#12\catcode `\^12\catcode `\_12\catcode `\%12\relax}%
\providecommand \@@startlink[1]{}%
\providecommand \@@endlink[0]{}%
\providecommand \url  [0]{\begingroup\@sanitize@url \@url }%
\providecommand \@url [1]{\endgroup\@href {#1}{\urlprefix }}%
\providecommand \urlprefix  [0]{URL }%
\providecommand \Eprint [0]{\href }%
\providecommand \doibase [0]{http://dx.doi.org/}%
\providecommand \selectlanguage [0]{\@gobble}%
\providecommand \bibinfo  [0]{\@secondoftwo}%
\providecommand \bibfield  [0]{\@secondoftwo}%
\providecommand \translation [1]{[#1]}%
\providecommand \BibitemOpen [0]{}%
\providecommand \bibitemStop [0]{}%
\providecommand \bibitemNoStop [0]{.\EOS\space}%
\providecommand \EOS [0]{\spacefactor3000\relax}%
\providecommand \BibitemShut  [1]{\csname bibitem#1\endcsname}%
\let\auto@bib@innerbib\@empty
%</preamble>
\bibitem [{\citenamefont {Andreev}(1964)}]{Andreev1964}%
  \BibitemOpen
  \bibfield  {author} {\bibinfo {author} {\bibfnamefont {A.~F.}\ \bibnamefont
  {Andreev}},\ }\href
  {http://www.jetp.ac.ru/cgi-bin/e/index/r/46/5/p1823?a=list} {\bibfield
  {journal} {\bibinfo  {journal} {ZhETF}\ }\textbf {\bibinfo {volume} {46}},\
  \bibinfo {pages} {1823} (\bibinfo {year} {1964})}\BibitemShut {NoStop}%
\bibitem [{\citenamefont {de~Jong}\ and\ \citenamefont
  {Beenakker}(1995)}]{Beenakker1995}%
  \BibitemOpen
  \bibfield  {author} {\bibinfo {author} {\bibfnamefont {M.~J.~M.}\
  \bibnamefont {de~Jong}}\ and\ \bibinfo {author} {\bibfnamefont {C.~W.~J.}\
  \bibnamefont {Beenakker}},\ }\href {\doibase 10.1103/PhysRevLett.74.1657}
  {\bibfield  {journal} {\bibinfo  {journal} {Phys. Rev. Lett.}\ }\textbf
  {\bibinfo {volume} {74}},\ \bibinfo {pages} {1657} (\bibinfo {year}
  {1995})}\BibitemShut {NoStop}%
\bibitem [{\citenamefont {Cayssol}\ and\ \citenamefont
  {Montambaux}(2005)}]{Cayssol2005}%
  \BibitemOpen
  \bibfield  {author} {\bibinfo {author} {\bibfnamefont {J.}~\bibnamefont
  {Cayssol}}\ and\ \bibinfo {author} {\bibfnamefont {G.}~\bibnamefont
  {Montambaux}},\ }\href {\doibase 10.1103/PhysRevB.71.012507} {\bibfield
  {journal} {\bibinfo  {journal} {Phys. Rev. B}\ }\textbf {\bibinfo {volume}
  {71}},\ \bibinfo {pages} {012507} (\bibinfo {year} {2005})}\BibitemShut
  {NoStop}%
\bibitem [{\citenamefont {Upadhyay}\ \emph {et~al.}(1998)\citenamefont
  {Upadhyay}, \citenamefont {Palanisami}, \citenamefont {Louie},\ and\
  \citenamefont {Buhrman}}]{Upadhyay1998}%
  \BibitemOpen
  \bibfield  {author} {\bibinfo {author} {\bibfnamefont {S.~K.}\ \bibnamefont
  {Upadhyay}}, \bibinfo {author} {\bibfnamefont {A.}~\bibnamefont
  {Palanisami}}, \bibinfo {author} {\bibfnamefont {R.~N.}\ \bibnamefont
  {Louie}}, \ and\ \bibinfo {author} {\bibfnamefont {R.~A.}\ \bibnamefont
  {Buhrman}},\ }\href {\doibase 10.1103/PhysRevLett.81.3247} {\bibfield
  {journal} {\bibinfo  {journal} {Phys. Rev. Lett.}\ }\textbf {\bibinfo
  {volume} {81}},\ \bibinfo {pages} {3247} (\bibinfo {year}
  {1998})}\BibitemShut {NoStop}%
\bibitem [{\citenamefont {Soulen}\ \emph {et~al.}(1999)\citenamefont {Soulen},
  \citenamefont {Osofsky}, \citenamefont {Nadgorny}, \citenamefont {Ambrose},
  \citenamefont {Broussard}, \citenamefont {Cheng}, \citenamefont {Byers},
  \citenamefont {Tanaka}, \citenamefont {Nowack}, \citenamefont {Moodera},
  \citenamefont {Laprade}, \citenamefont {Barry},\ and\ \citenamefont
  {Coey}}]{Soulen1999}%
  \BibitemOpen
  \bibfield  {author} {\bibinfo {author} {\bibfnamefont {R.~J.}\ \bibnamefont
  {Soulen}}, \bibinfo {author} {\bibfnamefont {M.~S.}\ \bibnamefont {Osofsky}},
  \bibinfo {author} {\bibfnamefont {B.}~\bibnamefont {Nadgorny}}, \bibinfo
  {author} {\bibfnamefont {T.}~\bibnamefont {Ambrose}}, \bibinfo {author}
  {\bibfnamefont {P.}~\bibnamefont {Broussard}}, \bibinfo {author}
  {\bibfnamefont {S.~F.}\ \bibnamefont {Cheng}}, \bibinfo {author}
  {\bibfnamefont {J.}~\bibnamefont {Byers}}, \bibinfo {author} {\bibfnamefont
  {C.~T.}\ \bibnamefont {Tanaka}}, \bibinfo {author} {\bibfnamefont
  {J.}~\bibnamefont {Nowack}}, \bibinfo {author} {\bibfnamefont {J.~S.}\
  \bibnamefont {Moodera}}, \bibinfo {author} {\bibfnamefont {G.}~\bibnamefont
  {Laprade}}, \bibinfo {author} {\bibfnamefont {A.}~\bibnamefont {Barry}}, \
  and\ \bibinfo {author} {\bibfnamefont {M.~D.}\ \bibnamefont {Coey}},\ }\href
  {\doibase 10.1063/1.370417} {\bibfield  {journal} {\bibinfo  {journal}
  {Journal of Applied Physics}\ }\textbf {\bibinfo {volume} {85}},\ \bibinfo
  {pages} {4589} (\bibinfo {year} {1999})},\ \Eprint
  {http://arxiv.org/abs/https://doi.org/10.1063/1.370417}
  {https://doi.org/10.1063/1.370417} \BibitemShut {NoStop}%
\bibitem [{\citenamefont {Haldane}(1988)}]{Haldane1988}%
  \BibitemOpen
  \bibfield  {author} {\bibinfo {author} {\bibfnamefont {F.~D.~M.}\
  \bibnamefont {Haldane}},\ }\href {\doibase 10.1103/PhysRevLett.61.2015}
  {\bibfield  {journal} {\bibinfo  {journal} {Phys. Rev. Lett.}\ }\textbf
  {\bibinfo {volume} {61}},\ \bibinfo {pages} {2015} (\bibinfo {year}
  {1988})}\BibitemShut {NoStop}%
\bibitem [{\citenamefont {Jotzu}\ \emph {et~al.}(2014)\citenamefont {Jotzu},
  \citenamefont {Messer}, \citenamefont {Desbuquois}, \citenamefont {Lebrat},
  \citenamefont {Uehlinger}, \citenamefont {Greif},\ and\ \citenamefont
  {Esslinger}}]{Jotzu2014}%
  \BibitemOpen
  \bibfield  {author} {\bibinfo {author} {\bibfnamefont {G.}~\bibnamefont
  {Jotzu}}, \bibinfo {author} {\bibfnamefont {M.}~\bibnamefont {Messer}},
  \bibinfo {author} {\bibfnamefont {R.}~\bibnamefont {Desbuquois}}, \bibinfo
  {author} {\bibfnamefont {M.}~\bibnamefont {Lebrat}}, \bibinfo {author}
  {\bibfnamefont {T.}~\bibnamefont {Uehlinger}}, \bibinfo {author}
  {\bibfnamefont {D.}~\bibnamefont {Greif}}, \ and\ \bibinfo {author}
  {\bibfnamefont {T.}~\bibnamefont {Esslinger}},\ }\href
  {http://dx.doi.org/10.1038/nature13915} {\bibfield  {journal} {\bibinfo
  {journal} {Nature}\ }\textbf {\bibinfo {volume} {515}},\ \bibinfo {pages}
  {237 EP } (\bibinfo {year} {2014})}\BibitemShut {NoStop}%
\bibitem [{\citenamefont {Rechtsman}\ \emph {et~al.}(2013)\citenamefont
  {Rechtsman}, \citenamefont {Zeuner}, \citenamefont {Plotnik}, \citenamefont
  {Lumer}, \citenamefont {Podolsky}, \citenamefont {Dreisow}, \citenamefont
  {Nolte}, \citenamefont {Segev},\ and\ \citenamefont
  {Szameit}}]{Rechtsman2013}%
  \BibitemOpen
  \bibfield  {author} {\bibinfo {author} {\bibfnamefont {M.~C.}\ \bibnamefont
  {Rechtsman}}, \bibinfo {author} {\bibfnamefont {J.~M.}\ \bibnamefont
  {Zeuner}}, \bibinfo {author} {\bibfnamefont {Y.}~\bibnamefont {Plotnik}},
  \bibinfo {author} {\bibfnamefont {Y.}~\bibnamefont {Lumer}}, \bibinfo
  {author} {\bibfnamefont {D.}~\bibnamefont {Podolsky}}, \bibinfo {author}
  {\bibfnamefont {F.}~\bibnamefont {Dreisow}}, \bibinfo {author} {\bibfnamefont
  {S.}~\bibnamefont {Nolte}}, \bibinfo {author} {\bibfnamefont
  {M.}~\bibnamefont {Segev}}, \ and\ \bibinfo {author} {\bibfnamefont
  {A.}~\bibnamefont {Szameit}},\ }\href {http://dx.doi.org/10.1038/nature12066}
  {\bibfield  {journal} {\bibinfo  {journal} {Nature}\ }\textbf {\bibinfo
  {volume} {496}},\ \bibinfo {pages} {196 EP } (\bibinfo {year}
  {2013})}\BibitemShut {NoStop}%
\bibitem [{\citenamefont {Goldman}\ \emph {et~al.}(2016)\citenamefont
  {Goldman}, \citenamefont {Budich},\ and\ \citenamefont
  {Zoller}}]{Goldman2016}%
  \BibitemOpen
  \bibfield  {author} {\bibinfo {author} {\bibfnamefont {N.}~\bibnamefont
  {Goldman}}, \bibinfo {author} {\bibfnamefont {J.~C.}\ \bibnamefont {Budich}},
  \ and\ \bibinfo {author} {\bibfnamefont {P.}~\bibnamefont {Zoller}},\ }\href
  {http://dx.doi.org/10.1038/nphys3803} {\bibfield  {journal} {\bibinfo
  {journal} {Nature Physics}\ }\textbf {\bibinfo {volume} {12}},\ \bibinfo
  {pages} {639 EP } (\bibinfo {year} {2016})}\BibitemShut {NoStop}%
\bibitem [{\citenamefont {Oka}\ and\ \citenamefont {Aoki}(2009)}]{Oka2009}%
  \BibitemOpen
  \bibfield  {author} {\bibinfo {author} {\bibfnamefont {T.}~\bibnamefont
  {Oka}}\ and\ \bibinfo {author} {\bibfnamefont {H.}~\bibnamefont {Aoki}},\
  }\href {\doibase 10.1103/PhysRevB.79.081406} {\bibfield  {journal} {\bibinfo
  {journal} {Phys. Rev. B}\ }\textbf {\bibinfo {volume} {79}},\ \bibinfo
  {pages} {081406} (\bibinfo {year} {2009})}\BibitemShut {NoStop}%
\bibitem [{\citenamefont {Yao}\ \emph {et~al.}(2007)\citenamefont {Yao},
  \citenamefont {MacDonald},\ and\ \citenamefont {Niu}}]{Yao2007}%
  \BibitemOpen
  \bibfield  {author} {\bibinfo {author} {\bibfnamefont {W.}~\bibnamefont
  {Yao}}, \bibinfo {author} {\bibfnamefont {A.~H.}\ \bibnamefont {MacDonald}},
  \ and\ \bibinfo {author} {\bibfnamefont {Q.}~\bibnamefont {Niu}},\ }\href
  {\doibase 10.1103/PhysRevLett.99.047401} {\bibfield  {journal} {\bibinfo
  {journal} {Phys. Rev. Lett.}\ }\textbf {\bibinfo {volume} {99}},\ \bibinfo
  {pages} {047401} (\bibinfo {year} {2007})}\BibitemShut {NoStop}%
\bibitem [{\citenamefont {Delplace}\ \emph {et~al.}(2013)\citenamefont
  {Delplace}, \citenamefont {G\'omez-Le\'on},\ and\ \citenamefont
  {Platero}}]{Delplace2013}%
  \BibitemOpen
  \bibfield  {author} {\bibinfo {author} {\bibfnamefont {P.}~\bibnamefont
  {Delplace}}, \bibinfo {author} {\bibfnamefont {A.}~\bibnamefont
  {G\'omez-Le\'on}}, \ and\ \bibinfo {author} {\bibfnamefont {G.}~\bibnamefont
  {Platero}},\ }\href {\doibase 10.1103/PhysRevB.88.245422} {\bibfield
  {journal} {\bibinfo  {journal} {Phys. Rev. B}\ }\textbf {\bibinfo {volume}
  {88}},\ \bibinfo {pages} {245422} (\bibinfo {year} {2013})}\BibitemShut
  {NoStop}%
\bibitem [{\citenamefont {Katan}\ and\ \citenamefont
  {Podolsky}(2013)}]{Katan2013}%
  \BibitemOpen
  \bibfield  {author} {\bibinfo {author} {\bibfnamefont {Y.~T.}\ \bibnamefont
  {Katan}}\ and\ \bibinfo {author} {\bibfnamefont {D.}~\bibnamefont
  {Podolsky}},\ }\href {\doibase 10.1103/PhysRevLett.110.016802} {\bibfield
  {journal} {\bibinfo  {journal} {Phys. Rev. Lett.}\ }\textbf {\bibinfo
  {volume} {110}},\ \bibinfo {pages} {016802} (\bibinfo {year}
  {2013})}\BibitemShut {NoStop}%
\bibitem [{\citenamefont {Lababidi}\ \emph {et~al.}(2014)\citenamefont
  {Lababidi}, \citenamefont {Satija},\ and\ \citenamefont
  {Zhao}}]{Lababidi2014}%
  \BibitemOpen
  \bibfield  {author} {\bibinfo {author} {\bibfnamefont {M.}~\bibnamefont
  {Lababidi}}, \bibinfo {author} {\bibfnamefont {I.~I.}\ \bibnamefont
  {Satija}}, \ and\ \bibinfo {author} {\bibfnamefont {E.}~\bibnamefont
  {Zhao}},\ }\href {\doibase 10.1103/PhysRevLett.112.026805} {\bibfield
  {journal} {\bibinfo  {journal} {Phys. Rev. Lett.}\ }\textbf {\bibinfo
  {volume} {112}},\ \bibinfo {pages} {026805} (\bibinfo {year}
  {2014})}\BibitemShut {NoStop}%
\bibitem [{\citenamefont {Iadecola}\ \emph {et~al.}(2013)\citenamefont
  {Iadecola}, \citenamefont {Campbell}, \citenamefont {Chamon}, \citenamefont
  {Hou}, \citenamefont {Jackiw}, \citenamefont {Pi},\ and\ \citenamefont
  {Kusminskiy}}]{Iadecola2013}%
  \BibitemOpen
  \bibfield  {author} {\bibinfo {author} {\bibfnamefont {T.}~\bibnamefont
  {Iadecola}}, \bibinfo {author} {\bibfnamefont {D.}~\bibnamefont {Campbell}},
  \bibinfo {author} {\bibfnamefont {C.}~\bibnamefont {Chamon}}, \bibinfo
  {author} {\bibfnamefont {C.-Y.}\ \bibnamefont {Hou}}, \bibinfo {author}
  {\bibfnamefont {R.}~\bibnamefont {Jackiw}}, \bibinfo {author} {\bibfnamefont
  {S.-Y.}\ \bibnamefont {Pi}}, \ and\ \bibinfo {author} {\bibfnamefont {S.~V.}\
  \bibnamefont {Kusminskiy}},\ }\href {\doibase 10.1103/PhysRevLett.110.176603}
  {\bibfield  {journal} {\bibinfo  {journal} {Phys. Rev. Lett.}\ }\textbf
  {\bibinfo {volume} {110}},\ \bibinfo {pages} {176603} (\bibinfo {year}
  {2013})}\BibitemShut {NoStop}%
\bibitem [{\citenamefont {Goldman}\ and\ \citenamefont
  {Dalibard}(2014)}]{Goldman2014}%
  \BibitemOpen
  \bibfield  {author} {\bibinfo {author} {\bibfnamefont {N.}~\bibnamefont
  {Goldman}}\ and\ \bibinfo {author} {\bibfnamefont {J.}~\bibnamefont
  {Dalibard}},\ }\href {\doibase 10.1103/PhysRevX.4.031027} {\bibfield
  {journal} {\bibinfo  {journal} {Phys. Rev. X}\ }\textbf {\bibinfo {volume}
  {4}},\ \bibinfo {pages} {031027} (\bibinfo {year} {2014})}\BibitemShut
  {NoStop}%
\bibitem [{\citenamefont {Grushin}\ \emph {et~al.}(2014)\citenamefont
  {Grushin}, \citenamefont {G\'omez-Le\'on},\ and\ \citenamefont
  {Neupert}}]{Grushin2014}%
  \BibitemOpen
  \bibfield  {author} {\bibinfo {author} {\bibfnamefont {A.~G.}\ \bibnamefont
  {Grushin}}, \bibinfo {author} {\bibfnamefont {A.}~\bibnamefont
  {G\'omez-Le\'on}}, \ and\ \bibinfo {author} {\bibfnamefont {T.}~\bibnamefont
  {Neupert}},\ }\href {\doibase 10.1103/PhysRevLett.112.156801} {\bibfield
  {journal} {\bibinfo  {journal} {Phys. Rev. Lett.}\ }\textbf {\bibinfo
  {volume} {112}},\ \bibinfo {pages} {156801} (\bibinfo {year}
  {2014})}\BibitemShut {NoStop}%
\bibitem [{\citenamefont {Dutreix}\ \emph {et~al.}(2016)\citenamefont
  {Dutreix}, \citenamefont {Stepanov},\ and\ \citenamefont
  {Katsnelson}}]{Dutreix2016}%
  \BibitemOpen
  \bibfield  {author} {\bibinfo {author} {\bibfnamefont {C.}~\bibnamefont
  {Dutreix}}, \bibinfo {author} {\bibfnamefont {E.~A.}\ \bibnamefont
  {Stepanov}}, \ and\ \bibinfo {author} {\bibfnamefont {M.~I.}\ \bibnamefont
  {Katsnelson}},\ }\href {\doibase 10.1103/PhysRevB.93.241404} {\bibfield
  {journal} {\bibinfo  {journal} {Phys. Rev. B}\ }\textbf {\bibinfo {volume}
  {93}},\ \bibinfo {pages} {241404} (\bibinfo {year} {2016})}\BibitemShut
  {NoStop}%
\bibitem [{\citenamefont {Stepanov}\ \emph {et~al.}(2017)\citenamefont
  {Stepanov}, \citenamefont {Dutreix},\ and\ \citenamefont
  {Katsnelson}}]{Stepanov2017}%
  \BibitemOpen
  \bibfield  {author} {\bibinfo {author} {\bibfnamefont {E.~A.}\ \bibnamefont
  {Stepanov}}, \bibinfo {author} {\bibfnamefont {C.}~\bibnamefont {Dutreix}}, \
  and\ \bibinfo {author} {\bibfnamefont {M.~I.}\ \bibnamefont {Katsnelson}},\
  }\href {\doibase 10.1103/PhysRevLett.118.157201} {\bibfield  {journal}
  {\bibinfo  {journal} {Phys. Rev. Lett.}\ }\textbf {\bibinfo {volume} {118}},\
  \bibinfo {pages} {157201} (\bibinfo {year} {2017})}\BibitemShut {NoStop}%
\bibitem [{\citenamefont {Kitagawa}\ \emph {et~al.}(2010)\citenamefont
  {Kitagawa}, \citenamefont {Berg}, \citenamefont {Rudner},\ and\ \citenamefont
  {Demler}}]{Kitagawa2010}%
  \BibitemOpen
  \bibfield  {author} {\bibinfo {author} {\bibfnamefont {T.}~\bibnamefont
  {Kitagawa}}, \bibinfo {author} {\bibfnamefont {E.}~\bibnamefont {Berg}},
  \bibinfo {author} {\bibfnamefont {M.}~\bibnamefont {Rudner}}, \ and\ \bibinfo
  {author} {\bibfnamefont {E.}~\bibnamefont {Demler}},\ }\href {\doibase
  10.1103/PhysRevB.82.235114} {\bibfield  {journal} {\bibinfo  {journal} {Phys.
  Rev. B}\ }\textbf {\bibinfo {volume} {82}},\ \bibinfo {pages} {235114}
  (\bibinfo {year} {2010})}\BibitemShut {NoStop}%
\bibitem [{\citenamefont {Lindner}\ \emph {et~al.}(2011)\citenamefont
  {Lindner}, \citenamefont {Refael},\ and\ \citenamefont
  {Galitski}}]{Lindner2011}%
  \BibitemOpen
  \bibfield  {author} {\bibinfo {author} {\bibfnamefont {N.~H.}\ \bibnamefont
  {Lindner}}, \bibinfo {author} {\bibfnamefont {G.}~\bibnamefont {Refael}}, \
  and\ \bibinfo {author} {\bibfnamefont {V.}~\bibnamefont {Galitski}},\ }\href
  {http://dx.doi.org/10.1038/nphys1926} {\bibfield  {journal} {\bibinfo
  {journal} {Nature Physics}\ }\textbf {\bibinfo {volume} {7}},\ \bibinfo
  {pages} {490 EP } (\bibinfo {year} {2011})},\ \bibinfo {note}
  {article}\BibitemShut {NoStop}%
\bibitem [{\citenamefont {D\'ora}\ \emph {et~al.}(2012)\citenamefont {D\'ora},
  \citenamefont {Cayssol}, \citenamefont {Simon},\ and\ \citenamefont
  {Moessner}}]{Dora2012}%
  \BibitemOpen
  \bibfield  {author} {\bibinfo {author} {\bibfnamefont {B.}~\bibnamefont
  {D\'ora}}, \bibinfo {author} {\bibfnamefont {J.}~\bibnamefont {Cayssol}},
  \bibinfo {author} {\bibfnamefont {F.}~\bibnamefont {Simon}}, \ and\ \bibinfo
  {author} {\bibfnamefont {R.}~\bibnamefont {Moessner}},\ }\href {\doibase
  10.1103/PhysRevLett.108.056602} {\bibfield  {journal} {\bibinfo  {journal}
  {Phys. Rev. Lett.}\ }\textbf {\bibinfo {volume} {108}},\ \bibinfo {pages}
  {056602} (\bibinfo {year} {2012})}\BibitemShut {NoStop}%
\bibitem [{\citenamefont {Kundu}\ \emph {et~al.}(2014)\citenamefont {Kundu},
  \citenamefont {Fertig},\ and\ \citenamefont {Seradjeh}}]{Kundu2014}%
  \BibitemOpen
  \bibfield  {author} {\bibinfo {author} {\bibfnamefont {A.}~\bibnamefont
  {Kundu}}, \bibinfo {author} {\bibfnamefont {H.~A.}\ \bibnamefont {Fertig}}, \
  and\ \bibinfo {author} {\bibfnamefont {B.}~\bibnamefont {Seradjeh}},\ }\href
  {\doibase 10.1103/PhysRevLett.113.236803} {\bibfield  {journal} {\bibinfo
  {journal} {Phys. Rev. Lett.}\ }\textbf {\bibinfo {volume} {113}},\ \bibinfo
  {pages} {236803} (\bibinfo {year} {2014})}\BibitemShut {NoStop}%
\bibitem [{\citenamefont {Titum}\ \emph {et~al.}(2015)\citenamefont {Titum},
  \citenamefont {Lindner}, \citenamefont {Rechtsman},\ and\ \citenamefont
  {Refael}}]{Titum2015}%
  \BibitemOpen
  \bibfield  {author} {\bibinfo {author} {\bibfnamefont {P.}~\bibnamefont
  {Titum}}, \bibinfo {author} {\bibfnamefont {N.~H.}\ \bibnamefont {Lindner}},
  \bibinfo {author} {\bibfnamefont {M.~C.}\ \bibnamefont {Rechtsman}}, \ and\
  \bibinfo {author} {\bibfnamefont {G.}~\bibnamefont {Refael}},\ }\href
  {\doibase 10.1103/PhysRevLett.114.056801} {\bibfield  {journal} {\bibinfo
  {journal} {Phys. Rev. Lett.}\ }\textbf {\bibinfo {volume} {114}},\ \bibinfo
  {pages} {056801} (\bibinfo {year} {2015})}\BibitemShut {NoStop}%
\bibitem [{\citenamefont {Klinovaja}\ \emph {et~al.}(2016)\citenamefont
  {Klinovaja}, \citenamefont {Stano},\ and\ \citenamefont
  {Loss}}]{Klinovaja2016}%
  \BibitemOpen
  \bibfield  {author} {\bibinfo {author} {\bibfnamefont {J.}~\bibnamefont
  {Klinovaja}}, \bibinfo {author} {\bibfnamefont {P.}~\bibnamefont {Stano}}, \
  and\ \bibinfo {author} {\bibfnamefont {D.}~\bibnamefont {Loss}},\ }\href
  {\doibase 10.1103/PhysRevLett.116.176401} {\bibfield  {journal} {\bibinfo
  {journal} {Phys. Rev. Lett.}\ }\textbf {\bibinfo {volume} {116}},\ \bibinfo
  {pages} {176401} (\bibinfo {year} {2016})}\BibitemShut {NoStop}%
\bibitem [{\citenamefont {Thakurathi}\ \emph {et~al.}(2013)\citenamefont
  {Thakurathi}, \citenamefont {Patel}, \citenamefont {Sen},\ and\ \citenamefont
  {Dutta}}]{Thakurathi2013}%
  \BibitemOpen
  \bibfield  {author} {\bibinfo {author} {\bibfnamefont {M.}~\bibnamefont
  {Thakurathi}}, \bibinfo {author} {\bibfnamefont {A.~A.}\ \bibnamefont
  {Patel}}, \bibinfo {author} {\bibfnamefont {D.}~\bibnamefont {Sen}}, \ and\
  \bibinfo {author} {\bibfnamefont {A.}~\bibnamefont {Dutta}},\ }\href
  {\doibase 10.1103/PhysRevB.88.155133} {\bibfield  {journal} {\bibinfo
  {journal} {Phys. Rev. B}\ }\textbf {\bibinfo {volume} {88}},\ \bibinfo
  {pages} {155133} (\bibinfo {year} {2013})}\BibitemShut {NoStop}%
\bibitem [{\citenamefont {Chen}\ and\ \citenamefont {Franz}(2016)}]{Chen2016}%
  \BibitemOpen
  \bibfield  {author} {\bibinfo {author} {\bibfnamefont {A.}~\bibnamefont
  {Chen}}\ and\ \bibinfo {author} {\bibfnamefont {M.}~\bibnamefont {Franz}},\
  }\href {\doibase 10.1103/PhysRevB.93.201105} {\bibfield  {journal} {\bibinfo
  {journal} {Phys. Rev. B}\ }\textbf {\bibinfo {volume} {93}},\ \bibinfo
  {pages} {201105} (\bibinfo {year} {2016})}\BibitemShut {NoStop}%
\bibitem [{\citenamefont {Khanna}\ \emph {et~al.}(2017)\citenamefont {Khanna},
  \citenamefont {Rao},\ and\ \citenamefont {Kundu}}]{Khanna2017}%
  \BibitemOpen
  \bibfield  {author} {\bibinfo {author} {\bibfnamefont {U.}~\bibnamefont
  {Khanna}}, \bibinfo {author} {\bibfnamefont {S.}~\bibnamefont {Rao}}, \ and\
  \bibinfo {author} {\bibfnamefont {A.}~\bibnamefont {Kundu}},\ }\href
  {\doibase 10.1103/PhysRevB.95.201115} {\bibfield  {journal} {\bibinfo
  {journal} {Phys. Rev. B}\ }\textbf {\bibinfo {volume} {95}},\ \bibinfo
  {pages} {201115} (\bibinfo {year} {2017})}\BibitemShut {NoStop}%
\bibitem [{\citenamefont {Aidelsburger}\ \emph {et~al.}(2014)\citenamefont
  {Aidelsburger}, \citenamefont {Lohse}, \citenamefont {Schweizer},
  \citenamefont {Atala}, \citenamefont {Barreiro}, \citenamefont
  {Nascimb{\`e}ne}, \citenamefont {Cooper}, \citenamefont {Bloch},\ and\
  \citenamefont {Goldman}}]{Aidelsburger2014}%
  \BibitemOpen
  \bibfield  {author} {\bibinfo {author} {\bibfnamefont {M.}~\bibnamefont
  {Aidelsburger}}, \bibinfo {author} {\bibfnamefont {M.}~\bibnamefont {Lohse}},
  \bibinfo {author} {\bibfnamefont {C.}~\bibnamefont {Schweizer}}, \bibinfo
  {author} {\bibfnamefont {M.}~\bibnamefont {Atala}}, \bibinfo {author}
  {\bibfnamefont {J.~.~T.}\ \bibnamefont {Barreiro}}, \bibinfo {author}
  {\bibfnamefont {S.}~\bibnamefont {Nascimb{\`e}ne}}, \bibinfo {author}
  {\bibfnamefont {N.~.~R.}\ \bibnamefont {Cooper}}, \bibinfo {author}
  {\bibfnamefont {I.}~\bibnamefont {Bloch}}, \ and\ \bibinfo {author}
  {\bibfnamefont {N.}~\bibnamefont {Goldman}},\ }\href
  {http://dx.doi.org/10.1038/nphys3171} {\bibfield  {journal} {\bibinfo
  {journal} {Nature Physics}\ }\textbf {\bibinfo {volume} {11}},\ \bibinfo
  {pages} {162 EP } (\bibinfo {year} {2014})}\BibitemShut {NoStop}%
\bibitem [{\citenamefont {Drummond}\ \emph {et~al.}(2012)\citenamefont
  {Drummond}, \citenamefont {Z\'olyomi},\ and\ \citenamefont
  {Fal'ko}}]{Drummond2012}%
  \BibitemOpen
  \bibfield  {author} {\bibinfo {author} {\bibfnamefont {N.~D.}\ \bibnamefont
  {Drummond}}, \bibinfo {author} {\bibfnamefont {V.}~\bibnamefont {Z\'olyomi}},
  \ and\ \bibinfo {author} {\bibfnamefont {V.~I.}\ \bibnamefont {Fal'ko}},\
  }\href {\doibase 10.1103/PhysRevB.85.075423} {\bibfield  {journal} {\bibinfo
  {journal} {Phys. Rev. B}\ }\textbf {\bibinfo {volume} {85}},\ \bibinfo
  {pages} {075423} (\bibinfo {year} {2012})}\BibitemShut {NoStop}%
\bibitem [{\citenamefont {Liu}\ \emph {et~al.}(2011{\natexlab{a}})\citenamefont
  {Liu}, \citenamefont {Feng},\ and\ \citenamefont {Yao}}]{Liu2011b}%
  \BibitemOpen
  \bibfield  {author} {\bibinfo {author} {\bibfnamefont {C.-C.}\ \bibnamefont
  {Liu}}, \bibinfo {author} {\bibfnamefont {W.}~\bibnamefont {Feng}}, \ and\
  \bibinfo {author} {\bibfnamefont {Y.}~\bibnamefont {Yao}},\ }\href {\doibase
  10.1103/PhysRevLett.107.076802} {\bibfield  {journal} {\bibinfo  {journal}
  {Phys. Rev. Lett.}\ }\textbf {\bibinfo {volume} {107}},\ \bibinfo {pages}
  {076802} (\bibinfo {year} {2011}{\natexlab{a}})}\BibitemShut {NoStop}%
\bibitem [{\citenamefont {Liu}\ \emph {et~al.}(2011{\natexlab{b}})\citenamefont
  {Liu}, \citenamefont {Jiang},\ and\ \citenamefont {Yao}}]{Liu2011a}%
  \BibitemOpen
  \bibfield  {author} {\bibinfo {author} {\bibfnamefont {C.-C.}\ \bibnamefont
  {Liu}}, \bibinfo {author} {\bibfnamefont {H.}~\bibnamefont {Jiang}}, \ and\
  \bibinfo {author} {\bibfnamefont {Y.}~\bibnamefont {Yao}},\ }\href {\doibase
  10.1103/PhysRevB.84.195430} {\bibfield  {journal} {\bibinfo  {journal} {Phys.
  Rev. B}\ }\textbf {\bibinfo {volume} {84}},\ \bibinfo {pages} {195430}
  (\bibinfo {year} {2011}{\natexlab{b}})}\BibitemShut {NoStop}%
\bibitem [{\citenamefont {Ezawa}(2013)}]{Ezawa2011a}%
  \BibitemOpen
  \bibfield  {author} {\bibinfo {author} {\bibfnamefont {M.}~\bibnamefont
  {Ezawa}},\ }\href {\doibase 10.1103/PhysRevLett.110.026603} {\bibfield
  {journal} {\bibinfo  {journal} {Phys. Rev. Lett.}\ }\textbf {\bibinfo
  {volume} {110}},\ \bibinfo {pages} {026603} (\bibinfo {year}
  {2013})}\BibitemShut {NoStop}%
\bibitem [{\citenamefont {Ezawa}(2015)}]{Ezawa2011b}%
  \BibitemOpen
  \bibfield  {author} {\bibinfo {author} {\bibfnamefont {M.}~\bibnamefont
  {Ezawa}},\ }\href {\doibase 10.7566/JPSJ.84.121003} {\bibfield  {journal}
  {\bibinfo  {journal} {Journal of the Physical Society of Japan}\ }\textbf
  {\bibinfo {volume} {84}},\ \bibinfo {pages} {121003} (\bibinfo {year}
  {2015})},\ \Eprint
  {http://arxiv.org/abs/https://doi.org/10.7566/JPSJ.84.121003}
  {https://doi.org/10.7566/JPSJ.84.121003} \BibitemShut {NoStop}%
\bibitem [{\citenamefont {Linder}\ and\ \citenamefont
  {Yokoyama}(2014)}]{Linder2014}%
  \BibitemOpen
  \bibfield  {author} {\bibinfo {author} {\bibfnamefont {J.}~\bibnamefont
  {Linder}}\ and\ \bibinfo {author} {\bibfnamefont {T.}~\bibnamefont
  {Yokoyama}},\ }\href {\doibase 10.1103/PhysRevB.89.020504} {\bibfield
  {journal} {\bibinfo  {journal} {Phys. Rev. B}\ }\textbf {\bibinfo {volume}
  {89}},\ \bibinfo {pages} {020504} (\bibinfo {year} {2014})}\BibitemShut
  {NoStop}%
\bibitem [{\citenamefont {Li}\ and\ \citenamefont {Zhang}(2016)}]{Li2016}%
  \BibitemOpen
  \bibfield  {author} {\bibinfo {author} {\bibfnamefont {K.}~\bibnamefont
  {Li}}\ and\ \bibinfo {author} {\bibfnamefont {Y.-Y.}\ \bibnamefont {Zhang}},\
  }\href {\doibase 10.1103/PhysRevB.94.165441} {\bibfield  {journal} {\bibinfo
  {journal} {Phys. Rev. B}\ }\textbf {\bibinfo {volume} {94}},\ \bibinfo
  {pages} {165441} (\bibinfo {year} {2016})}\BibitemShut {NoStop}%
\bibitem [{\citenamefont {Qiu}\ \emph {et~al.}(2016)\citenamefont {Qiu},
  \citenamefont {Cheng}, \citenamefont {Cao},\ and\ \citenamefont
  {Qin}}]{Qiu2016}%
  \BibitemOpen
  \bibfield  {author} {\bibinfo {author} {\bibfnamefont {X.~J.}\ \bibnamefont
  {Qiu}}, \bibinfo {author} {\bibfnamefont {Y.~F.}\ \bibnamefont {Cheng}},
  \bibinfo {author} {\bibfnamefont {Z.~Z.}\ \bibnamefont {Cao}}, \ and\
  \bibinfo {author} {\bibfnamefont {C.~C.}\ \bibnamefont {Qin}},\ }\href
  {http://stacks.iop.org/0022-3727/49/i=24/a=245101} {\bibfield  {journal}
  {\bibinfo  {journal} {Journal of Physics D: Applied Physics}\ }\textbf
  {\bibinfo {volume} {49}},\ \bibinfo {pages} {245101} (\bibinfo {year}
  {2016})}\BibitemShut {NoStop}%
\bibitem [{\citenamefont {Wang}\ \emph {et~al.}(2015)\citenamefont {Wang},
  \citenamefont {Hao},\ and\ \citenamefont {Chan}}]{Wang2015}%
  \BibitemOpen
  \bibfield  {author} {\bibinfo {author} {\bibfnamefont {J.}~\bibnamefont
  {Wang}}, \bibinfo {author} {\bibfnamefont {L.}~\bibnamefont {Hao}}, \ and\
  \bibinfo {author} {\bibfnamefont {K.~S.}\ \bibnamefont {Chan}},\ }\href
  {\doibase 10.1103/PhysRevB.91.085415} {\bibfield  {journal} {\bibinfo
  {journal} {Phys. Rev. B}\ }\textbf {\bibinfo {volume} {91}},\ \bibinfo
  {pages} {085415} (\bibinfo {year} {2015})}\BibitemShut {NoStop}%
\bibitem [{\citenamefont {Chen}\ \emph {et~al.}(2013)\citenamefont {Chen},
  \citenamefont {Feng},\ and\ \citenamefont {Wu}}]{Chen2013}%
  \BibitemOpen
  \bibfield  {author} {\bibinfo {author} {\bibfnamefont {L.}~\bibnamefont
  {Chen}}, \bibinfo {author} {\bibfnamefont {B.}~\bibnamefont {Feng}}, \ and\
  \bibinfo {author} {\bibfnamefont {K.}~\bibnamefont {Wu}},\ }\href {\doibase
  10.1063/1.4793998} {\bibfield  {journal} {\bibinfo  {journal} {Applied
  Physics Letters}\ }\textbf {\bibinfo {volume} {102}},\ \bibinfo {pages}
  {081602} (\bibinfo {year} {2013})},\ \Eprint
  {http://arxiv.org/abs/https://doi.org/10.1063/1.4793998}
  {https://doi.org/10.1063/1.4793998} \BibitemShut {NoStop}%
\bibitem [{\citenamefont {Vosoughi-nia}\ \emph {et~al.}(2017)\citenamefont
  {Vosoughi-nia}, \citenamefont {Hajati},\ and\ \citenamefont
  {Rashedi}}]{Vosoughi-nia2017}%
  \BibitemOpen
  \bibfield  {author} {\bibinfo {author} {\bibfnamefont {S.}~\bibnamefont
  {Vosoughi-nia}}, \bibinfo {author} {\bibfnamefont {Y.}~\bibnamefont
  {Hajati}}, \ and\ \bibinfo {author} {\bibfnamefont {G.}~\bibnamefont
  {Rashedi}},\ }\href {\doibase 10.1063/1.4996347} {\bibfield  {journal}
  {\bibinfo  {journal} {Journal of Applied Physics}\ }\textbf {\bibinfo
  {volume} {122}},\ \bibinfo {pages} {043906} (\bibinfo {year} {2017})},\
  \Eprint {http://arxiv.org/abs/https://doi.org/10.1063/1.4996347}
  {https://doi.org/10.1063/1.4996347} \BibitemShut {NoStop}%
\bibitem [{\citenamefont {Zhou}\ and\ \citenamefont {Jin}(2016)}]{Zhou2016}%
  \BibitemOpen
  \bibfield  {author} {\bibinfo {author} {\bibfnamefont {X.}~\bibnamefont
  {Zhou}}\ and\ \bibinfo {author} {\bibfnamefont {G.}~\bibnamefont {Jin}},\
  }\href {\doibase 10.1103/PhysRevB.94.165436} {\bibfield  {journal} {\bibinfo
  {journal} {Phys. Rev. B}\ }\textbf {\bibinfo {volume} {94}},\ \bibinfo
  {pages} {165436} (\bibinfo {year} {2016})}\BibitemShut {NoStop}%
\bibitem [{\citenamefont {Zhou}\ and\ \citenamefont {Jin}(2017)}]{Zhou2017}%
  \BibitemOpen
  \bibfield  {author} {\bibinfo {author} {\bibfnamefont {X.}~\bibnamefont
  {Zhou}}\ and\ \bibinfo {author} {\bibfnamefont {G.}~\bibnamefont {Jin}},\
  }\href {\doibase 10.1103/PhysRevB.95.195419} {\bibfield  {journal} {\bibinfo
  {journal} {Phys. Rev. B}\ }\textbf {\bibinfo {volume} {95}},\ \bibinfo
  {pages} {195419} (\bibinfo {year} {2017})}\BibitemShut {NoStop}%
\bibitem [{\citenamefont {Dolcini}\ and\ \citenamefont
  {Dell'Anna}(2008)}]{Dolcini2008}%
  \BibitemOpen
  \bibfield  {author} {\bibinfo {author} {\bibfnamefont {F.}~\bibnamefont
  {Dolcini}}\ and\ \bibinfo {author} {\bibfnamefont {L.}~\bibnamefont
  {Dell'Anna}},\ }\href {\doibase 10.1103/PhysRevB.78.024518} {\bibfield
  {journal} {\bibinfo  {journal} {Phys. Rev. B}\ }\textbf {\bibinfo {volume}
  {78}},\ \bibinfo {pages} {024518} (\bibinfo {year} {2008})}\BibitemShut
  {NoStop}%
\bibitem [{\citenamefont {Liu}\ \emph {et~al.}(2011{\natexlab{c}})\citenamefont
  {Liu}, \citenamefont {Jiang},\ and\ \citenamefont {Yao}}]{Liu2011}%
  \BibitemOpen
  \bibfield  {author} {\bibinfo {author} {\bibfnamefont {C.-C.}\ \bibnamefont
  {Liu}}, \bibinfo {author} {\bibfnamefont {H.}~\bibnamefont {Jiang}}, \ and\
  \bibinfo {author} {\bibfnamefont {Y.}~\bibnamefont {Yao}},\ }\href {\doibase
  10.1103/PhysRevB.84.195430} {\bibfield  {journal} {\bibinfo  {journal} {Phys.
  Rev. B}\ }\textbf {\bibinfo {volume} {84}},\ \bibinfo {pages} {195430}
  (\bibinfo {year} {2011}{\natexlab{c}})}\BibitemShut {NoStop}%
\bibitem [{\citenamefont {Konschuh}\ \emph {et~al.}(2010)\citenamefont
  {Konschuh}, \citenamefont {Gmitra},\ and\ \citenamefont
  {Fabian}}]{Konschuh2010}%
  \BibitemOpen
  \bibfield  {author} {\bibinfo {author} {\bibfnamefont {S.}~\bibnamefont
  {Konschuh}}, \bibinfo {author} {\bibfnamefont {M.}~\bibnamefont {Gmitra}}, \
  and\ \bibinfo {author} {\bibfnamefont {J.}~\bibnamefont {Fabian}},\ }\href
  {\doibase 10.1103/PhysRevB.82.245412} {\bibfield  {journal} {\bibinfo
  {journal} {Phys. Rev. B}\ }\textbf {\bibinfo {volume} {82}},\ \bibinfo
  {pages} {245412} (\bibinfo {year} {2010})}\BibitemShut {NoStop}%
\bibitem [{\citenamefont {Mohan}\ \emph {et~al.}(2016)\citenamefont {Mohan},
  \citenamefont {Saxena}, \citenamefont {Kundu},\ and\ \citenamefont
  {Rao}}]{Mohan2016}%
  \BibitemOpen
  \bibfield  {author} {\bibinfo {author} {\bibfnamefont {P.}~\bibnamefont
  {Mohan}}, \bibinfo {author} {\bibfnamefont {R.}~\bibnamefont {Saxena}},
  \bibinfo {author} {\bibfnamefont {A.}~\bibnamefont {Kundu}}, \ and\ \bibinfo
  {author} {\bibfnamefont {S.}~\bibnamefont {Rao}},\ }\href {\doibase
  10.1103/PhysRevB.94.235419} {\bibfield  {journal} {\bibinfo  {journal} {Phys.
  Rev. B}\ }\textbf {\bibinfo {volume} {94}},\ \bibinfo {pages} {235419}
  (\bibinfo {year} {2016})}\BibitemShut {NoStop}%
\bibitem [{\citenamefont {Beenakker}(2006)}]{Beenakker2006}%
  \BibitemOpen
  \bibfield  {author} {\bibinfo {author} {\bibfnamefont {C.~W.~J.}\
  \bibnamefont {Beenakker}},\ }\href {\doibase 10.1103/PhysRevLett.97.067007}
  {\bibfield  {journal} {\bibinfo  {journal} {Phys. Rev. Lett.}\ }\textbf
  {\bibinfo {volume} {97}},\ \bibinfo {pages} {067007} (\bibinfo {year}
  {2006})}\BibitemShut {NoStop}%
\bibitem [{\citenamefont {Bukov}\ \emph {et~al.}(2015)\citenamefont {Bukov},
  \citenamefont {D'Alessio},\ and\ \citenamefont
  {Polkovnikov}}]{Polkovnikov2015}%
  \BibitemOpen
  \bibfield  {author} {\bibinfo {author} {\bibfnamefont {M.}~\bibnamefont
  {Bukov}}, \bibinfo {author} {\bibfnamefont {L.}~\bibnamefont {D'Alessio}}, \
  and\ \bibinfo {author} {\bibfnamefont {A.}~\bibnamefont {Polkovnikov}},\
  }\href {\doibase 10.1080/00018732.2015.1055918} {\bibfield  {journal}
  {\bibinfo  {journal} {Advances in Physics}\ }\textbf {\bibinfo {volume}
  {64}},\ \bibinfo {pages} {139} (\bibinfo {year} {2015})},\ \Eprint
  {http://arxiv.org/abs/https://doi.org/10.1080/00018732.2015.1055918}
  {https://doi.org/10.1080/00018732.2015.1055918} \BibitemShut {NoStop}%
\bibitem [{\citenamefont {Maricq}(1982)}]{Maricq1982}%
  \BibitemOpen
  \bibfield  {author} {\bibinfo {author} {\bibfnamefont {M.~M.}\ \bibnamefont
  {Maricq}},\ }\href {\doibase 10.1103/PhysRevB.25.6622} {\bibfield  {journal}
  {\bibinfo  {journal} {Phys. Rev. B}\ }\textbf {\bibinfo {volume} {25}},\
  \bibinfo {pages} {6622} (\bibinfo {year} {1982})}\BibitemShut {NoStop}%
\bibitem [{\citenamefont {Dehghani}\ \emph {et~al.}(2014)\citenamefont
  {Dehghani}, \citenamefont {Oka},\ and\ \citenamefont {Mitra}}]{Mitra2014}%
  \BibitemOpen
  \bibfield  {author} {\bibinfo {author} {\bibfnamefont {H.}~\bibnamefont
  {Dehghani}}, \bibinfo {author} {\bibfnamefont {T.}~\bibnamefont {Oka}}, \
  and\ \bibinfo {author} {\bibfnamefont {A.}~\bibnamefont {Mitra}},\ }\href
  {\doibase 10.1103/PhysRevB.90.195429} {\bibfield  {journal} {\bibinfo
  {journal} {Phys. Rev. B}\ }\textbf {\bibinfo {volume} {90}},\ \bibinfo
  {pages} {195429} (\bibinfo {year} {2014})}\BibitemShut {NoStop}%
\bibitem [{\citenamefont {Mori}\ \emph {et~al.}(2016)\citenamefont {Mori},
  \citenamefont {Kuwahara},\ and\ \citenamefont {Saito}}]{Mori2016}%
  \BibitemOpen
  \bibfield  {author} {\bibinfo {author} {\bibfnamefont {T.}~\bibnamefont
  {Mori}}, \bibinfo {author} {\bibfnamefont {T.}~\bibnamefont {Kuwahara}}, \
  and\ \bibinfo {author} {\bibfnamefont {K.}~\bibnamefont {Saito}},\ }\href
  {\doibase 10.1103/PhysRevLett.116.120401} {\bibfield  {journal} {\bibinfo
  {journal} {Phys. Rev. Lett.}\ }\textbf {\bibinfo {volume} {116}},\ \bibinfo
  {pages} {120401} (\bibinfo {year} {2016})}\BibitemShut {NoStop}%
\end{thebibliography}%


\begin{thebibliography}{10}
% --- Floquet Band Structure

\bibitem{Khanna_2017}
U.~Khanna, S.~Rao, and A.~Kundu,



%--- High frequency
\bibitem{Polkovnikov2015}
M.~Bukov, L~D'Alessio, and A.~Polkovnikov,
Advances in Physics, {\bf 64}, 2  (2015).


\bibitem{Mohan}
P. Mohan, R. Saxena, A. Kundu and S. Rao,
\newblock Phys. Rev. B {\bf 94}, 235419 (2016).


%-- Floquet Spins
%\bibitem{FlSpin1}
%M.~Bukov, M.~Kolodrubetz, and A.~Polkovnikov,
%Phys. Rev. Lett. {\bf 116}, 125301 (2016).

%\bibitem{FlSpin2}
%M.~Claassen, H.-C.~Jiang, B.~Moritz, and T.~P.~Devereaux,
%arXiv:1611.07964 (2016).

%\bibitem{FlSpin3}
%S.~Kitamura, T.~Oka, and H.~Aoki,
%Phys. Rev. B {\bf 96}, 014406 (2017).

% --- Floquet Experiments --------------




%---------------------------------



% --- AR in Si..
\bibitem{Fabrizio}
Fabrizio Dolcini and Luca Dell’Anna


\bibitem{Linder_si}
J.~Linder and T.~Yokoyama,
Phys. Rev. B {\bf 89}, 020504(R) (2014).




\bibitem{subgap_sil2}
S.~Vosoughi-nia, Y.~Hajati, and G.~Rashedi,
Journal of Applied Physics {\bf122}, 043906 (2017).

%-- OTHER
\bibitem{Konschuh_2010}
S.~Konschuh, M.~Gmitra, and J.~Fabian,
\newblock \prb\ {\bf 82}, 245412 (2010).




% Heating
\bibitem{Maricq}
M. M. Maricq, Phys. Rev. B {\bf 25}, 6622 (1982).

\bibitem{Mitra}
H. Dehghani, T. Oka, and A. Mitra, Phys. Rev. B {\bf 90}, 195429 (2014).

\bibitem{Mori}
T. Mori, T. Kuwahara, and Keiji Saito, Phys. Rev. Lett. {\bf 116}, 120401 (2016)

\end{thebibliography}

\iffalse

\fi

\end{document}